\documentclass[preprint,authoryear,sort&compress,12pt]{elsarticle}  
\usepackage[top=1.25in, bottom=1.25in, left=1.25in, right=1.25in]{geometry}
\usepackage{graphicx}
\usepackage{amssymb}
\usepackage{amsmath}
\usepackage{lineno}
\usepackage{subfig}

\usepackage{color}
\usepackage{soul}
\usepackage{array}
\usepackage{hyperref}
\usepackage[colorinlistoftodos]{todonotes}

\newcommand{\xmb}[1]{\ensuremath{\mathbf{#1}}}

\newcommand*\patchAmsMathEnvironmentForLineno[1]{%
  \expandafter\let\csname old#1\expandafter\endcsname\csname #1\endcsname
  \expandafter\let\csname oldend#1\expandafter\endcsname\csname end#1\endcsname
  \renewenvironment{#1}%
     {\linenomath\csname old#1\endcsname}%
     {\csname oldend#1\endcsname\endlinenomath}}%
\newcommand*\patchBothAmsMathEnvironmentsForLineno[1]{%
  \patchAmsMathEnvironmentForLineno{#1}%
  \patchAmsMathEnvironmentForLineno{#1*}}%
\AtBeginDocument{%
\patchBothAmsMathEnvironmentsForLineno{equation}%
\patchBothAmsMathEnvironmentsForLineno{align}%
\patchBothAmsMathEnvironmentsForLineno{flalign}%
\patchBothAmsMathEnvironmentsForLineno{alignat}%
\patchBothAmsMathEnvironmentsForLineno{gather}%
\patchBothAmsMathEnvironmentsForLineno{multline}%
}

\journal{International Journal of Multiphase Flow}

\begin{document}

\begin{frontmatter}

 \title{Diffusion-Based Coarse Graining in Hybrid Continuum--Discrete Solvers: Theoretical Formulation and A Priori Tests}

 \author{Rui Sun} \ead{sunrui@vt.edu}
 \author{Heng Xiao\corref{corxh}} \ead{hengxiao@vt.edu}
 \address{Department of Aerospace and Ocean Engineering, Virginia Tech, Blacksburg, VA 24060, United
 States}

 \cortext[corxh]{Corresponding author. Tel: +1 540 231 0926}

\begin{abstract}
  Coarse graining is an important ingredient in many multi-scale continuum-discrete solvers such as CFD--DEM (computational fluid dynamics--discrete element method) solvers for dense particle-laden flows. Although CFD--DEM solvers have become a mature technique that is widely used in multiphase flow research and industrial flow simulations, a flexible and easy-to-implement coarse graining algorithm that can work with CFD solvers of arbitrary meshes is still lacking. In this work, we proposed a new coarse graining algorithm for continuum--discrete solvers for dense particle-laden flows based on solving a transient diffusion equation. Via theoretical analysis we demonstrated that the proposed method is equivalent to the statistical kernel method with a Gaussian kernel, but the current method is much more straightforward to implement in CFD--DEM solvers. \textit{A priori} numerical tests were performed to obtain the solid volume fraction fields based on given particle distributions, the results obtained by using the proposed algorithm were compared with those from other coarse graining methods in the literature (e.g., the particle centroid method, the divided particle volume method, and the two-grid formulation).  The numerical tests demonstrated that the proposed coarse graining procedure based on solving diffusion equations is theoretically sound, easy to implement and parallelize in general CFD solvers, and has improved mesh-convergence characteristics compared with existing coarse graining methods. The diffusion-based coarse graining method has been implemented into a CFD--DEM solver, the results of which are presented in a separate work (R. Sun and H. Xiao, Diffusion-based coarse graining in hybrid continuum-discrete solvers: Application in CFD-DEM solvers for particle laden flows.  International Journal of Multiphase Flow 72, 233–247).
\end{abstract}


\end{frontmatter}


\section{Introduction}
\label{sec:intro}

\subsection{Coarse Graining in Continuum--Discrete Modeling}

In computational mechanics for solids and fluids, continuum methods based on partial differential
equations, e.g., the Navier-Stokes or elasticity equations, are usually discretized by the
finite-difference, finite-volume, or finite-element method and used to investigate macroscopic
system responses, e.g., structural deformations or fluid
flows~\citep{mitchell80FDM,versteeg07FVM,zienkiewicz71FEM}. On the other hand, discrete methods such
as molecular dynamics and direct simulation Monte Carlo methods are used to simulate microscopic
properties of systems at length- and time-scales that are much smaller than those studies by using
continuum methods~\citep{piekos96nm,belytschko02as}. Continuum and discrete methods are
complementary to each other, not only in terms of the length- and time-scales they cover, but also
because they are used for different purposes. 

Traditionally, continuum and discrete methods have been developed in separate communities without
significant interactions. This is partly a reflection of the scale separation in classical continuum
mechanics; that is, the scales of the representative volume element (RVE) are many orders of
magnitude larger than those of the individual molecules or atoms therein.  As a consequence, the
phenomena of interests in the macroscopic scales and those in the microscopic scales are
dramatically different. However, the past few decades have seen a surge of interests in the
development of methods aiming for bridging the continuum and the microscopic scales. These efforts
originate from several communities with a diverse physical settings ranging from fracture
mechanics~\citep{belytschko03coup,xiao04ab} and complex fluids~\citep{Donev10AH} to materials
sciences~\citep[e.g.,][]{Delgado05,eidel09CA,muller13sp,rottler13ma,curtin13mf}. In spite of these
efforts, many theoretical and practical challenges exist in these multi-scale models. A particular
difficulty that is common among most continuum--discrete solvers is the coupling between the
continuum solver and the discrete solver with guaranteed conservation of relevant quantities. This
is due to the fact that the conservation equations on the two scales are formulated for quantities
of vastly different natures, and thus the bridging between the microscopic and macroscopic
quantities are critical yet extremely difficult to achieve~\citep{xiao04ab,Delgado05}. 
In this work we highlight the difficulties of coarse graining in the context of continuum--discrete
solvers for dense particle-laden multiphase flows. Subsequently, a strategy that is theoretically
meritorious and convenient for numerical implementations is proposed.  However, since the
difficulties associated with coarse graining are common among many other continuum--discrete
methods, it is expected that the strategy developed in this work shall be useful for those methods
as well.

\subsection{Coarse Graining in CFD--DEM Solvers}

Dense particle-laden flows are encountered in a wide range of industrial applications and natural
processes including fluidized-bed reactors in chemical engineering~\citep{muller08gr, muller09va},
pneumatic conveyors in the mineral processing industries~\citep{han03DEM}, and sediment transport
processes in riverine and coastal flows~\citep{drake01dp, calantoni04ms}. A class of
continuum--discrete methods have been proposed and developed in the past two decades, where the
fluid (the carrier phase) is modeled with a continuum approach in the Eulerian framework, while the
particles (the dispersed phase) are tracked individually in the Lagrangian framework accounting for
 the inter-particle collisions and fluid--particle interactions. Compared with other alternatives
such as the two-fluid model and direct numerical simulations~\citep{Kempe_12_CM,Kempe:14on,
  esmaeeli98DNS1, esmaeeli99DNS2}, the continuum--discrete approach is able to resolve much of the
particle flow physics without requiring excessive computational costs. As such, it has received much
attention in multiphase flow research communities and gained increasing popularity in industrial
applications as well.

 Discrete Element Method (DEM) is introduced by \cite{cundall79} and is widely used in
  the modeling of dry granular flows where the fluid flow is not dynamically important (except for
  cohesion and lubrication forces).  Readers are referred to a recent review by
  \cite{guo2015discrete} on this subject.  Proposed in the 1990s by \cite{tsuji93}, the CFD
(Computational Fluid Dynamics)--DEM approach is the earliest developed continuum--discrete method
for dense particle-laden flows. Other variants such as LES (Large Eddy Simulation)--DEM and SPH
(Smooth Particle Hydrodynamics)--DEM have been developed
recently~\citep[e.g.,][]{Zhou2004,potapov2001liquid,sun2013three}. However, since the coarse
graining procedure is common among this type of the continuum-discrete methods, in this work we
describe the coarse graining method with CFD--DEM applications in mind. The proposed method is,
however, applicable to other similar methods as well.

In CFD--DEM, the fluid is described by the volume-averaged Navier--Stokes
equations~\citep{anderson67}.  The effects of the particle-to-fluid interactions are accounted for
in the fluid continuity and momentum equations mainly through the presence of the following three
terms: solid volume fraction \(\varepsilon_s\), and solid phase velocities \(\mathbf{U}_s\),
fluid--particle forces \(\mathbf{F}_{fp}\) (primarily consisting of drag force $\mathbf{F}_{d}$ and
buoyancy). These are field quantities based on the CFD mesh, and are obtained from the locations,
the sizes, and the velocities of each individual particles as well as the fluid forces acting on
them.  {\color{black} The mapping from particle-scale quantities to macroscopic quantities is also
  referred to as coarse graining, averaging, or aggregation in the literature~\citep{xiao-cicp,
    zhu02ave}. We will use ``coarse graining'' and ``averaging'' interchangeably in this
  work.}\footnote{\color{black} The terminology ``coarse graining'' here should not be confused with
  its usage in other contexts. In DEM (and in molecular dynamics) simulations, ``coarse-graining''
  is also used to denote the process of representing a large number of real particles with a small
  number of ``super-particles'' to reduce computational costs while retaining essential dynamics of
  the system.} This seemingly simple operation involves several challenges when performed in
practical CFD--DEM solvers.  First, theoretically the procedure is not unique~\citep{zhu02ave}, but
the coarse grained fields can have profound influences on the CFD--DEM simulation results. Second,
there are several constraints on the procedure, e.g., it must conserve the relevant physical
quantities such as particle mass and momentum, both for the interior particles and for those near
the domain boundaries (e.g., walls).  Satisfying these constraints simultaneously is
challenging. Moreover, since industrial CFD simulations often involve complex geometries and
necessitate using meshes of poor quality, the procedure should be able to accommodate these meshes
without negatively impacting the CFD--DEM simulation results. Finally, it should be easy to
implement and amiable to parallelization, a technology that is ubiquitously used in modern CFD
codes. Numerous averaging techniques have been proposed and used in the
literature~\citep{wu09accu,wu09three,zhu02ave,xiao-cicp}.  To the authors' knowledge, however, a
method satisfying all the criteria above is still lacking. The objective of this work is to develop
an averaging strategy that is theoretically sound, easy to implement in industrial CFD solvers, and
capable of handling generic meshes used in these CFD solvers.  Due to space considerations, the
current paper focuses primarily on the theoretical analysis and \textit{a priori} evaluations of the
proposed method without actually testing it in a CFD--DEM solver. Its implementation in CFD--DEM
solvers and the evaluation of the performances are deferred to a separate, companion paper of the
present work~\citep{Xiao-IJMF}.

The rest of the paper is organized as follows.  In Section~\ref{sec:cg-summary} the desirable
features of averaging methods are discussed in detail, and existing algorithms in the
literature are reviewed, compared, and evaluated against the properties in this
list. Section~\ref{sec:cg} presents the proposed averaging algorithm, demonstrates its
theoretical equivalence to the statistical kernel method, and examines its characteristics based on
the theoretical analysis.  In Section~\ref{sec:apriori}, \textit{a priori} averaging tests are
conducted to examine the performance of the proposed method and to compare with the results obtained
with other averaging methods. Implementation details and computational costs considerations
are discussed in Section~\ref{sec:discuss}. The paper is concluded in Section~\ref{sec:conclude}.

\section{Review of Existing Coarse Graining Methods}
\label{sec:cg-summary}
Some of the commonly used and recently developed coarse graining methods in the CFD--DEM literature
include the particle centroid method (PCM), the divided particle volume method (DPVM), the
statistical kernel method, and the recently proposed two-grid formulation. These methods will be
critically reviewed below with their advantages and shortcomings examined. The discussion here is
constrained to particle-resolving simulations where particle--fluid interfaces are not resolved. The
continuum--discrete solvers where particle interfaces are explicitly resolved, e.g., with immersed
boundary method~\citep{Kempe:14on} or Lattice-Boltzmann method~\citep{chen91la,yin08,yin2009fluid},
are beyond the scope of the present work, since the averaging algorithms discussed in this
paper are not directly applicable to those methods. The methods where particles are represented as
porous bodies are omitted from the discussion here as well.

\subsection{Desirable Properties of Coarse Graining Procedure in CFD--DEM Solvers}
\label{sec:wish}
In the context of implementation in CFD--DEM solvers for particle-laden flows, we first outline
below a few desirable properties that a coarse graining method should have. This ``wish list'' will
serve as the basis for the evaluation and comparison of the existing and proposed coarse graining
methods.

Based on physical reasoning and from our experiences in conducting CFD--DEM simulations, a coarse
graining procedure should ideally:
\begin{enumerate}
\item
 conserve relevant physical quantities.

 For example, when calculating solid volume fraction field \( \varepsilon_s \), the particle phase
 mass should be conserved, i.e., the total mass computed from the coarse grained continuum field
 should be the same as the total particle mass in the discrete phase.   The conservation requirement
 can be written as follows:
 \begin{equation}
  \label{eq:conserve}
   \rho_s\sum_{k = 1}^{N_c} \varepsilon_{s, k} \, V_{c, k}
   = \sum_{i = 1}^{N_p} \rho_s \, V_{p, i} \; ,
 \end{equation}
 where the density \( \rho_s \) is assumed to be constant for all particles; \(N_c\) is the number
 of cells in the CFD mesh or the mesh used for averaging, which are sometimes the same;
 \(N_p\) is the number of particles in the system; \(\varepsilon_{s, k}\) and \(V_{c, k}\) are the
 solid volume fraction and the volume, respectively, of cell $k$.  Similar conservation requirements
 must be met for the momentum of the particles and that of the entire fluid--particle system as
 well. That is, when calculating solid phase velocity, the total momentum of the particles should be
 conserved; to conserve momentum in the fluid--particle system the total particle forces on the
 fluid should have the same magnitude as the sum of the forces on all particles but with opposite
 direction.

\item
 be able to handle particles both in the interior cells and the cells near boundaries without
 producing artifacts, including physical boundaries and processor boundaries in parallel computing.
\item
be able to achieve relatively mesh-independent results;

The mesh independence or mesh convergence basically requires that the mesh used in computational
simulations should be fine enough so that further mesh refinements do not lead to changes in the
results. While this is a basic requirement for almost any mesh-based discretization methods for
partial differential equations (e.g., finite difference, finite element), mesh convergences are not
trivial to achieve in CFD--DEM simulations. The reason is that in most existing averaging
methods the mesh is not only used for the discretization of the CFD conservation equations, but also
determines the coarse graining bandwidth, an important computational parameter. Similar difficulties
with mesh-convergence are well-known for LES with implicit filtering, where the mesh cell size
determines the filter size, which is a computational parameter~\citep{sagaut02LES}.
\item
 be convenient for implementation in parallel, three-dimensional CFD solvers based on unstructured
 meshes with arbitrary cell shapes;
\item
 be able to produce smooth coarse grained fields even with the presence of a few large particles in
 relatively small cells.

 In a simulation using a uniform CFD mesh and a particle system with a wide particle diameter
 distribution (e.g., well-sorted natural sand), the cell size to particle diameter ratio can be
 small at the vicinity of the large particles. Moreover, in practical CFD--DEM simulations small
 cells are often required in certain regions to resolve the flow features therein (e.g., shear
 layers or jets with strong velocity gradients~\citep{pope00tf}),
 particularly when high fidelity flow solvers such as those based on LES are used for the fluid
 phase. It is important for the averaging algorithms to be able to handle these situations,
 since abnormally large values or gradients in the solid volume fraction, velocity, or
 fluid--particle interaction forces, even in only a few cells, can destabilize the entire simulation
 or cause unphysical artifacts.
\end{enumerate}
Items 1 and 2 above are essential, without which the averaging procedure would cause the
CFD--DEM solvers to produce unphysical results. Items 3--5 are highly desirable, without which the
averaging procedure would impose significant restrictions on the CFD--DEM solvers and limit
their applications in practical simulations.

\subsection{Particle Centroid Method}
\label{sec:pcm}
In the particle centroid method, only the particles whose centroids fall within a cell are counted
for calculating the coarse grained continuum quantities (e.g., solid volume fraction
\(\varepsilon_s\), solid phase velocity \(\mathbf{U}_s\), drag force \(\mathbf{F}_d\)) for this
cell. For simplicity we use here the calculation of \(\varepsilon_s\) to illustrate the difference
among the averaging methods. The averaging of other variables such as solid phase
velocity and drag forces can be performed in the same way. 
In the configuration illustrated in Fig.~\ref{fig:cgsum}(a), with the PCM all
the particles, P1, P2, and P3, are counted towards the solid volume fraction of cell 2. All other
cells have zero solid volume fraction. It can be seen that when the particle centroid is located
near cell boundaries (e.g., particle P2 in Fig.~\ref{fig:cgsum}(a)), the error can be up to
50\%~\citep{peng14in}. If a cell contains several such particles and if these particles are large
compared to the host cell, unphysically large values can occur for this cell, leading to a
non-smooth solid particle volume field with large gradients. Such non-smoothness in the coarse
grained fields often causes spurious features in CFD--DEM simulations. However, except for this
drawback, this seemingly unsophisticated method works very well otherwise, particularly when the
cells are large compared to particle diameters. Moreover, if done properly the PCM can be made to
satisfy the conservation requirement specified above in a straightforward manner. Finally, its
implementation is straightforward even in parallel CFD solvers with unstructured meshes. As such,
the PCM is one of the most widely used averaging procedures in the
literature~\citep[e.g.,][]{zhu03te}.

\subsection{Divided Particle Volume Method}
\label{sec:dpv} 
\begin{figure}[!htbp]
 \centering \includegraphics[width=0.72\textwidth]{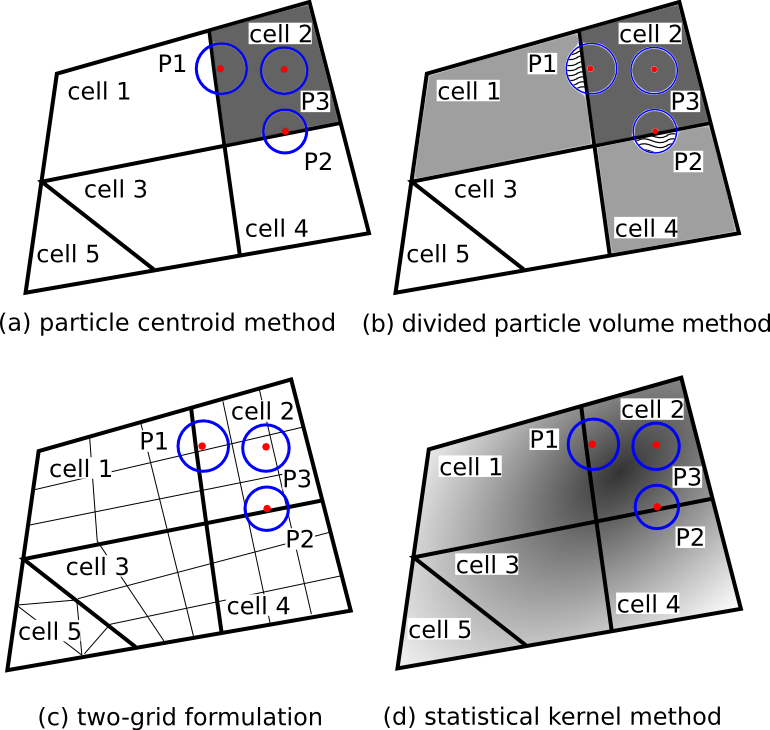}
 \caption{ 
  \label{fig:cgsum}
  Schematic of the four coarse graining methods: (a) the particle centroid method (PCM), (b) the
  divided particle volume method (DPVM), (c) the two-grid formulation, and (d) the statistical
  kernel function method.  In (a), (b), and (d), the shades indicate the coarse
  grained solid volume fraction fields \(\varepsilon_s\). In (a) the PCM, \(\varepsilon_s\) is
  nonzero only for cell 2, which contains the centroids of all particles; in (b) the DPVM, all cells
  1, 2, and 4, which overlap with a portion of at least one particle have nonzero \(\varepsilon_s\);
  in (c) two grid formulation, two separate meshes are used for coarse graining (thick lines) and
  fluid simulation (thin lines); in (d) the statistical kernel method, theoretically all cells can
  have non-zero \(\varepsilon_s\).}
\end{figure}

The divided particle volume method was proposed by
  \cite{wu09accu,wu09three} to address the above-mentioned drawbacks of the particle centroid
  method caused by particles intersecting multiple cells. It is also referred to as analytical
  method in the literature~\citep{peng14in}.  Instead of counting the entire particle volume to the
cell containing its centroid, in DPVM a particle's volume is divided among all cells it overlaps,
with each cell receiving the actual volume inside it. This is illustrated in
Fig.~\ref{fig:cgsum}(b). The volume of particle P1 is divided between cells 1 and 2, with cell 1
sharing the shaded portion. The volume of particle P2 is similarly shared between cells 4 and 2.  It
has been demonstrated~\citep{peng14in} that the dividing particle volume method gives much smoother
coarse grained solid volume fraction field than the PCM does, and the corresponding CFD--DEM results
were significantly improved as well. However, in unstructured CFD meshes where cells can be of
arbitrary shape, dividing particle volumes among intersecting cells requires accounting for many
scenarios, which is tedious to implement.  Even for a three-dimensional Cartesian mesh with cubic
cells (the simplest mesh one can hope for in practical CFD simulations), there are several
particle--cell intersecting scenarios to account for.  Theoretically, DPVM works for arbitrary
meshes, structured or unstructured, with any elements shapes as long as any edge of the cell has a
length larger than the particle diameter.  While \cite{wu09accu,wu09three}
  implemented a DPVM for typical cells shapes (e.g., tetrahedra and wedge) based on a commercial
  finite volume CFD solver, a DPVM implementation that is general enough to accounts for all cell
  shapes (e.g., an arbitrary combination of general polyhedra) is a daunting challenge for most CFD
  practitioners. We are not aware of any such implementations beyond those by Wu and his
  co-workers~\citep{wu09accu,wu09three,wu14parallel,peng14in}. Moreover, if the mesh contains small
cells with sizes comparable to or smaller than some large particles, the analytical method would
still lead to unphysically large solid volume fractions in these small cells.

\subsection{Two-Grid Formulation}
\label{sec:two-grid}
Another recent attempt to address the deficiency of the PCM in the cases of small cell-size to
particle diameter ratios was made by \cite{Deb13-ano}. Their idea was to use a separate mesh for the
averaging and another mesh for the CFD simulation.  The two meshes are constructed
independently. Specifically, the averaging mesh is chosen based on particle diameters to ensure that
the cell sizes \(\Delta x\) are larger than particle diameters \(d_p\) (they used \(\Delta x =
3d_p\)), and the CFD mesh is chosen according to the flow resolution requirements. {\color{black} They
  used structured Cartesian meshes similar to the one shown in Fig.~\ref{fig:diff-tg}(a), and
  Fig.~\ref{fig:cgsum} represents our interpretation of how it will be generalized to unstructured
  meshes.} As can be seen from Fig.~\ref{fig:diff-tg}(c), each cell in the coarse graining mesh
(displayed in thicker lines), contains exactly \(3 \times 3\) fluid mesh cells (thin lines). This
configuration with aligned cell boundaries simplifies the implementation dramatically, and it is
relatively easy to achieve on a structured Cartesian mesh. In parallel CFD solvers with unstructured
meshes, however, the implementation of the two-grid formulation is not straightforward.  For these
solvers, using two meshes with arbitrarily aligned cells would increase the complexity of the
implementation and necessitate non-trivial interpolations between the two meshes.  Alternatively, to
ensure alignment of averaging mesh cells and CFD cells, the latter of which may be of arbitrary
shape, one can construct a coarsened mesh based on a fine CFD mesh via agglomeration. In fact, mesh
agglomeration is common in geometric multi-grid methods for linear equation
systems~\citep{nishikawa11de}, which is used in many general-purpose unstructured CFD solvers. Such
a fine CFD mesh and the corresponding agglomerated coarse mesh is shown schematically in
Fig.~\ref{fig:cgsum}(c). However, agglomeration across processor boundaries in parallel computing is
not easy to implement, and in multi-grid methods for linear system the agglomeration is generally
not performed across cells on different processors~\citep{nishikawa11de}. Restricting agglomeration
within the mesh on the same processor would impose serious limitations on the effectiveness of the
two-grid formulation. Recently, \cite{su15two} proposed a dual-mesh method CFD--DEM with
conservation of mass and momentum guaranteed, but they did not discuss issues related to parallel
implementations. \cite{jing15extended}, on the other hand, used porous sphere model to handle large
particles in the system to avoid unphysically large volume fractions.

\subsection{Statistical Kernel Method}
\label{sec:kernel}
Yet another class of averaging methods are based on statistical kernel functions, which will
receive a more thorough discussion here since it is the foundation for the method  proposed in this
work.
\subsubsection{Formulation for Particles in Interior Cells}
\label{sec:kernel-interior}
 In the statistical kernel method the volume of each particle is distributed to the entire
domain according to a weight function called kernel function \(h(\mathbf{x})\). This is illustrated
in Fig.~\ref{fig:cgsum}(d). The solid volume fraction at location \(\mathbf{x}\) consists of the
superposition of the distributed volumes from all particles, i.e.,
\begin{equation}
 \varepsilon(\mathbf{x}) = \sum_{i=1}^{N_p} V_{p, i } h_i
 \label{eq:superpo}
\end{equation}
where \(h_i = h(\mathbf{x}-\mathbf{x}_i)\) is the weight function for particle \(i\). Possible
choices of kernel functions include top hat function (with which the method essentially degenerates
to PCM), Gaussian distribution function~\citep{xiao-cicp,glasser01scale}, and Johnson's \(S_B\)
distribution function~\citep{johnson1949,zhu02ave}. The Gaussian kernel function is of particular
interests to the current study due to its analytical tractability. In three-dimensional space the
Gaussian kernel function for particle $i$ is:
\begin{equation}
 h_i = h(\mathbf{x}-\mathbf{x}_i) = \frac{1}{(b^2 {\pi})^{3/2}} 
\exp \left[ - \frac{(\mathbf{x}-\mathbf{x}_i)^T (\mathbf{x}-\mathbf{x}_i) }{b^2} \right]
 \label{eq:hi}
\end{equation}
where superscript $T$ indicates vector transpose; \(b\) is the bandwidth of the Gaussian kernel,
indicating the size of the region influenced by the particle (i.e., the cells whose solid volume
fractions ``receives'' contribution from the particle); \(\mathbf{x}_i\) is the location of particle
\(i\). Note that for the averaging procedure to have the conservation property, the kernel
function \(h\) must satisfy the normalization condition, i.e.,
\begin{equation}
 \int_{\mathbb{R}^3} h(\mathbf{x}) \, d\mathbf{x} = 1,
 \label{eq:normal}
\end{equation}
where \(\mathbb{R}^3\) is the entire three-dimensional spatial domain. The Gaussian distribution
function as in Eq.~(\ref{eq:hi}) is known to satisfy this condition. As pointed out by
\cite{zhu02ave}, in addition to the normalization requirement, a kernel function should have local,
finite, and compact support, i.e., it is nonzero only at a finite region in the neighborhood of the
peak. Although the Gaussian distribution function has an infinite support theoretically, over 99\%
of the weight distribution is contained within a sphere of radius $3b$, which can be considered as
the effective support of the kernel. In numerical codes such as CFD--DEM solvers, the Gaussian
kernel can be considered as a function with effective local support~\citep{sun09}.

\begin{figure}[!htbp]
 \centering \includegraphics[width=0.4\textwidth]{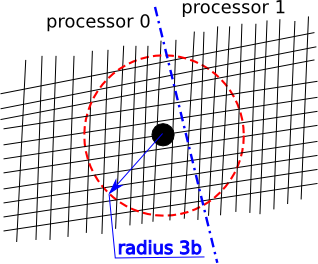}
 \caption{Distribution of particle volume to cells in the statistical kernel method.  The non-local
   nature of the method is illustrated by the fact that the cells whose solid volume fractions are
   influenced by the particle (indicated as those inside the dashed circle) can include
   and non-local cells far away from the particle, possibly on different processor (processor 1)
   than the one on which the particle is located (processor 0).}
 \label{fig:influence}
\end{figure}

Statistical kernel based averaging is a theoretically elegant method for coarse graining from the
discrete particle phase to a continuum descriptions of granular flows~\citep{zhu02ave,babic97avg}. It
resembles the technique used to bridge molecular dynamics in micro-scales and continuum mechanics in
macro-scales. \cite{zhu02ave} demonstrated that with the statistical averaging one can
derive from the equations at the DEM level to a continuum level balance equation for the granular
flow. Posterior tests showed good performance of the scheme by using the Johnson's $S_B$
distribution function.  However, its implementation in CFD--DEM solvers is not trivial since the
volume of a particle are distributed not only to local cells, but also to nonlocal cells that may be
located many layers of cells away from the particle location (e.g., those inside the dashed circle
in Fig.~\ref{fig:influence}), some of which may be even located on another processor. Searching for
these cells in unstructured meshes is challenging, since the stored connectivity information in
unstructured meshes normally only allows each cell's immediate neighbors to be found. To find
information of cells further away, one needs to find the immediate neighbors, and then find the
neighbors of all immediate neighbors, and repeat this process recursively until all the cells
``receiving'' the volume of the particle are identified~\citep{xiao-cicp}. This process needs to be
repeated for the volume distribution of each particle. The recursive nature of these search
operations makes them time-consuming, although it is performed only once in the beginning of the
simulation if the mesh does not move during the simulation. More importantly, the need for particles
to communicate to non-local cells makes it challenging for implementations in parallel solvers based
on MPI (Message Passing Interface)~\citep{Walker96mpi}, a technology that is utilized by most modern
CFD solvers~\citep[e.g.,][]{duggleby11}.

\subsubsection{Treatment of Physical Boundaries (Walls)}
\label{sec:kernel-withbnd}
\begin{figure}[!htbp]
 \centering \includegraphics[width=0.35\textwidth]{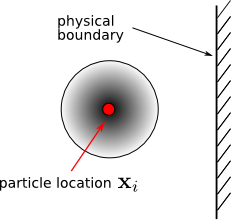}
 \caption{ 
  \label{fig:kernel-phy}
  Distribution of particle volume to cells in the statistical kernel method for interior particles
  away from boundaries.  The shade shows the distributed volume of the particle with the kernel
  function \(h(\mathbf{x}-\mathbf{x}_i)\).  The solid volume fraction of the cells outside the
  circle (regions that are not shaded) are not influenced by this particle, and thus the
  presence of the physical boundary does not influence its volume distribution.}
\end{figure}

The scheme in Eq.~(\ref{eq:hi}) is applicable for particles in the interior cells. That is,
particles that are ``far away'' from wall boundaries, i.e., those whose influence regions do not
insect with boundaries. This scenario is depicted in Fig.~\ref{fig:kernel-phy}.  Here we are only
concerned with physical boundaries of the system, i.e., mostly wall boundaries, although symmetry
plane boundaries can be treated in the same way. Computational boundaries such as inter-processor
boundaries are not considered since they should not pose theoretical difficulties.

\begin{figure}[!htbp]
 \centering \includegraphics[width=0.7\textwidth]{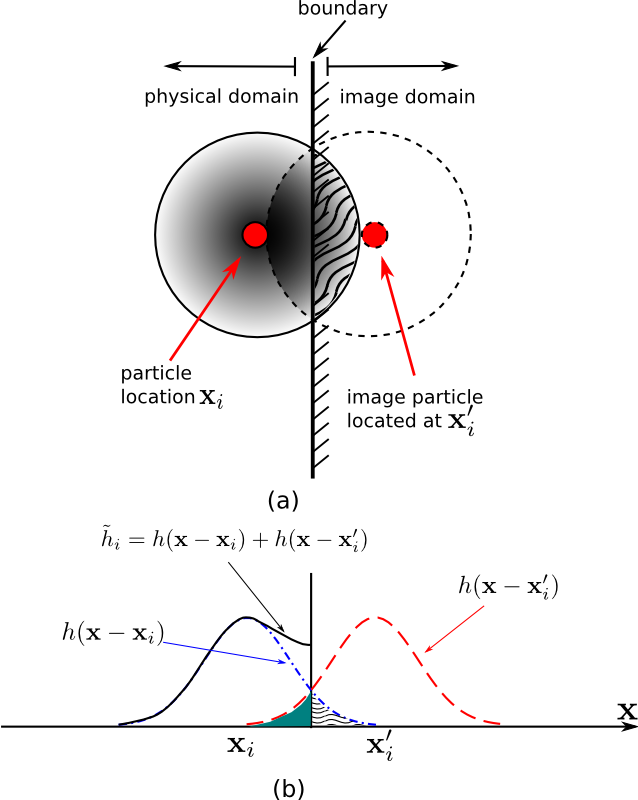}
 \caption{Statistical kernel based averaging for a particle volume near a physical boundary.
  The kernel function \(h(\mathbf{x}-\mathbf{x}_i)\) for the physical particle located at
  \(\mathbf{x}_i\) is shown on the physical domain (left to the boundary); the kernel function \(h(
  \mathbf{x}-\mathbf{x}'_i) \) for the image particle located at \(\mathbf{x}'_i\) is shown on the
  image domain (right to the boundary). The kernel function \(\tilde{h}_i =
  h(\mathbf{x}-\mathbf{x}_i) + h(\mathbf{x}-\mathbf{x}'_i)\) based on the superposition of the two
  functions above are shown on the physical domain in the top panel with shade and in the bottom
  panel with solid line.
  \label{fig:kernel-img}
 }
\end{figure}

For particles that are located near wall boundaries, \cite{zhu02ave} modified the kernel
function \(h_i\) in Eq.~(\ref{eq:hi}) to the following:
\begin{equation}
 \tilde{h}_i = h(\mathbf{x} - \mathbf{x}_i) + h(\mathbf{x} - \mathbf{x}'_i) ,
 \label{eq:hi-bound}
\end{equation}
where \(\mathbf{x}'_i\) is the location of the image of the particle located at \(\mathbf{x}_i\)
with respect to the physical boundary, shown in Fig.~\ref{fig:kernel-img}(b). They showed that the
modified kernel function \(\tilde{h}_i\) also satisfies the normalization condition in
Eq.~(\ref{eq:normal}). This modification has a straightforward physical interpretation as
illustrated in Fig.~\ref{fig:kernel-img}. When a particle is close to the physical boundary, if the
kernel function \(h_i = h(\mathbf{x} - \mathbf{x}_i)\) as in Eq.~(\ref{eq:hi}) were used, part of the
volume (indicated as wavy pattern in Figs.~\ref{fig:kernel-img}(a) and~\ref{fig:kernel-img}(b)) would
be distributed to regions outside the physical domain. In view of the normalization condition
Eq.~(\ref{eq:normal}), this would violate the conservation principle, since the integration domain
does not cover the entire space.  It can be seen that the kernel function \(h(\mathbf{x} -
\mathbf{x}'_i)\) adds the same amount that is distributed outside the physical domain by the kernel
function \(h_i\) to inside the domain (the shaded region in Fig.~\ref{fig:kernel-img}(b)), exactly
compensating for the lost part and thus restoring the conservation.  Although other methods such as
renormalization can also restore the conservation~\citep{xiao-cicp,ries14CG}, the image kernel method
is clearly more physical and elegant.

\begin{figure}[!htbp]
  \subfloat[][particle near orthogonal boundaries]{
    \centering \includegraphics[width=0.45\textwidth]{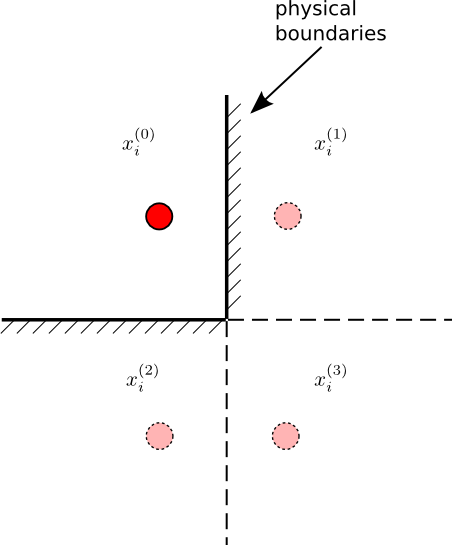} 
  }
  \subfloat[][particle near non-orthogonal boundaries]{
    \centering \includegraphics[width=0.45\textwidth]{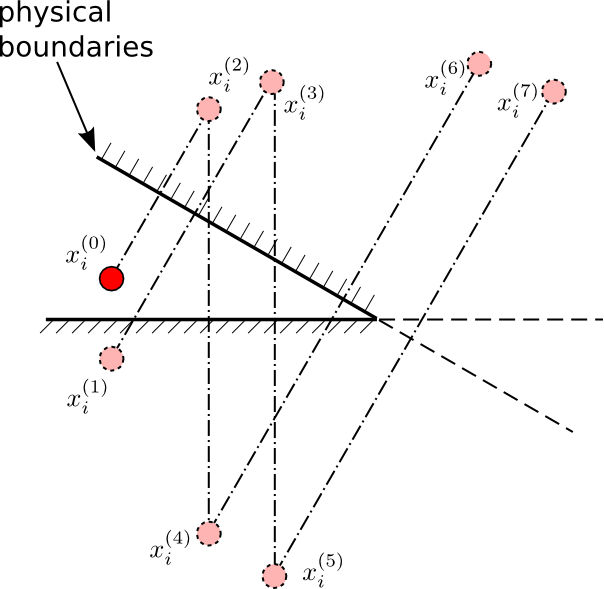} 
  }
 \caption{
  \label{fig:wall-ext}
  Physical and image kernels for a particle located near multiple boundaries showing the scenarios
  of (a) orthogonal and (b) non-orthogonal boundaries, the later of which involve the summation an
  infinite series of kernels and only the first eight terms are shown here. }
\end{figure}

Although not explicitly treated by \cite{zhu02ave}, the method of images above as in
Eq.~\ref{eq:hi-bound} can be extended to treat more complex boundaries. The most straightforward
extension is to account for the present of two boundaries, e.g., a particle located at the corner in
a two-dimensional domain, as illustrated in Fig.~\ref{fig:wall-ext}(a). In this case the modified
kernel function consists of four terms, with one associated with the physical particle and three
with image particles. The kernel function for particle $i$ is thus
\begin{equation}
  \label{eq:2dwall-ext}
  \tilde{h}_i = \sum_{n = 0}^{n=3} h^{(n)}_i \; ,
\end{equation}
where $n$ is the index of the terms in the series, with $h^{(0)}$ being the physical kernel and
other being image kernels. For a three-dimensional corner, the total number of terms in the kernel
function would be eight, with one physical kernel and seven image kernels. Graphical illustration
for this case is omitted here. For all the three cases above, the modified kernel functions satisfy
the normalization requirement exactly regardless of the kernel bandwidths and the distances between
the particle and the boundaries, thanks to the orthogonality of the boundaries in the two- and
three-dimensional cases.  For a particle located near multiple boundaries that are non-orthogonal, a
scenario that can be encountered in complex geometries, the exact expression for the modified kernel
function consists of an infinite series:
\begin{equation}
  \label{eq:wall-ext}
  \tilde{h}_i = \sum_{n = 0}^{\infty} h^{(n)}_i \; ,
\end{equation}
where $n$ is the index of the individual kernels, with $h^{(0)}_i$ centered at the physical particle
and $h^{(n)}_i$ (where $n = 1, 2, \cdots, \infty$) centered at the image particles.  This is
illustrated in Fig.~\ref{fig:wall-ext}(b), showing the locations of the physical kernel and the
first seven image kernels. Since the kernels have finite support (bandwidth), the image kernels in
the series located far away outside the computational domain can be safely truncated without
impairing the accuracy. In spite of this, it can be seen that the complexity of the kernel function
associated with a near-boundary particle, particularly for one that is located near a corner with
several non-orthogonal boundaries, is significant. Indeed, averaging strategy near boundaries
is an important and difficult subject that has been the focus of several recent
publications~\citep{ries14CG,weinhart12FD}. On the other hand, we point out that, at least in the
context of CFD--DEM simulations, this difficulty related to boundaries is unique to sophisticated
averaging schemes such as the statistical kernel method, as the presence of boundaries hardly
poses any difficulties for the relatively simple averaging schemes such as PCM and DPVM,
although in some cases they can make it difficult to construct a coarse graining mesh in the
two-grid formulation.

\subsection{Overview of Proposed Diffusion-Based Coarse Graining Method}
\label{sec:current}
In this work, we propose a diffusion-based coarse graining method consisting of the following two
steps (using again the solid volume fraction \(\varepsilon_s\) as example):
\begin{enumerate}
\item
 Perform averaging based on the particle centroid method to obtain the coarse grained field, denote
 the field as \(\varepsilon_{0}\)); and
\item Integrate a transient, homogeneous diffusion equation for a period of pseudo-time \(T\), using
  the field \( \varepsilon_{0} \) as initial condition and no-flux conditions on all physical
  boundaries. The obtained field \( \varepsilon_s (T) \) is the coarse grained field.
\end{enumerate}

In the rest of the paper we will show that this scheme is theoretically equivalent to the
statistical coarse graining method presented in Section~\ref{sec:kernel}, and the pseudo time-span
\(T\) is determined according to the bandwidth of the kernel function. The equivalence is valid even
with the presence of boundaries, and no special treatment for boundaries is needed in the proposed
method.  We will further demonstrate the following merits of the proposed method:
\begin{enumerate}
\item
 Straightforward implementation in almost any CFD code by solving a few additional diffusion
 equations;
\item
 Ability to obtain smooth coarse grained fields even with cell sizes comparable or slightly smaller
 than particle diameters; and
\item
 Relatively mesh-independent; and
\item
 No additional difficulties in implementations in parallel CFD solver using unstructured mesh with
 arbitrary cell shapes.
\end{enumerate}
The advantages and limitations of the averaging methods reviewed above are summarized in
Table~\ref{tab:pro-con}.

\begin{table}[!htbp]
  \caption{Comparison of advantages and disadvantages of coarse graining methods in the literature
  and the currently proposed method.}
 \begin{center}
   \begin{tabular}{>{\centering\arraybackslash}m{3cm}|>{\centering\arraybackslash}m{3.2cm}>{\centering\arraybackslash}m{2.3cm}
       >{\centering\arraybackslash}m{2.2cm} >{\centering\arraybackslash}m{2cm}}
   \hline coarse-graining methods &
   implementation in generic CFD solvers$^{(a)}$ &
   smoothness for small $\Delta x/d_p$$^{(b)}$ &
   mesh convergence & 
 treatment of {\color{black}  physical} boundaries \\
  \hline PCM &
     easy &
     poor &
     poor &
     easy \\
   DPVM &
     difficult &
     moderate\(^{(c)}\) &
     poor &
     easy \\
   statistical kernel  &
     moderate &
     good &
     good$^{(d)}$ &
     difficult \\
   two-grid &
     difficult &
     moderate &
     good$^{(d)}$ &
     moderate \\
   current method &
     easy &
{ \color{black} good } &
     good$^{(d)}$ &
     easy \\
   \hline
  \end{tabular}
 \end{center}
\small
(a) a parallel CFD solver based on unstructured mesh with cells of arbitrary shapes and
connectivity\\
(b) \(\Delta x / d_p \) indicates cell-size to particle diameter ratio, where \(\Delta x\) is mesh
cell size, and \(d_p\) is particle diameter \\
(c) if \(\Delta x / d_p \) is approximately 2 or larger \\
(d) if the kernel bandwidth (or coarse graining mesh size) is chosen independent of the CFD mesh
cell size
 \label{tab:pro-con}
\end{table}

\section{Diffusion-Based Coarse Graining Algorithm}
\label{sec:cg}

\subsection{Summary and Heuristic Reasoning}
\label{sec:diff-alg}
In the particle centroid method, the averaging from Lagrangian particle data to Eulerian
mesh-based field (e.g., solid volume fraction) is done as follows. We loop through all cells, and
for each cell we sum up all the particles volume to their host cells (defined as the cell within
which the particle centroid is located) to obtain the total particle volume in each cell. The solid
volume fraction of each cell \(k\) is then obtained by dividing the total particle volume in the
cell by the total volume of the cell \(V_{c, k}\).
That is,
\begin{equation}
 \varepsilon_{s,k} 
 = \frac{\sum_{i=1}^{n_{p, k}} V_{p, i}}{V_{c, k}} \, , \label{eq:pcm-k}
\end{equation}
where \(n_{p, k}\) is the number of particles in cell \(k\), which implies that \(\sum_{k=1}^{N_c}
n_{p, k} = N_p\). Multiplying both sides of Eq.~(\ref{eq:pcm-k}) by $V_{c, k}$ and taking summation
over all cells, the conservation requirements in Eq.~(\ref{eq:conserve}) can be
recovered. Therefore, the PCM-based averaging is conservative by construction.

As mentioned in Section~\ref{sec:pcm}, the coarse grained fields so obtained may have large
gradients. The cells with large solid volume fraction are shown schematically as three box-car
functions in Fig.~\ref{fig:diffNp}. The ticks indicate cell centers. To address this issue of
non-smoothness, a straightforward idea is to smooth this field via diffusion, i.e., by solving a
diffusion with non-smooth fields as initial conditions. The diffusion equation essentially
redistributes particle volumes within the field while automatically conserving total solid volume
in the domain during the diffusion. To ensure conservation of mass, no-flux (i.e., zero gradient)
conditions should be specified at all physical boundaries except for periodic boundaries, where
periodic conditions should be used instead. The smoothened field so obtained is shown in
Fig.~\ref{fig:diffNp} as thick dashed line.

\subsection{Equivalence to Statistical Kernel Method}
\label{sec:connect}

\subsubsection{Particles in Interior Cells}
Consider the diffusion equation for \(\varepsilon(\mathbf{x}, \tau)\) in the three-dimensional free
space:
\begin{align}
 \frac{\partial \varepsilon}{\partial \tau} &
 =  \nabla^2 \varepsilon \quad \textrm{ for } \; \mathbf{x} \in \mathbb{R}^3, \tau > 0
 \label{eq:diffusion}
 \\
 \varepsilon(\mathbf{x}, 0) &
 = \varepsilon_0 (\mathbf{x}) \label{eq:diffusion-b}
\end{align}
where \(\mathbf{x} \equiv [x, y, z]^T\) are spatial coordinates; \( \varepsilon \) can be considered
as solid volume fraction $\varepsilon_s$, but the derivations below are valid for other
coarse-grained quantities (e.g., \( \xmb{U}_s \) and \( \xmb{F}_s\)) as well.  Diffusion equations
are independently solved for each component when the coarse grained fields are vector- and
tensor-fields. For simplicity, the subscript $s$ of \( \varepsilon \) is omitted in this
subsection. \(\nabla^2 \varepsilon= \frac{\partial^2 \varepsilon}{\partial x^2} + \frac{\partial^2
  \varepsilon}{\partial y^2} + \frac{\partial^2 \varepsilon}{\partial z^2}\) in the Cartesian
coordinate; \( \varepsilon_0 (\mathbf{x})\) is a square integrable function; \(\tau\) is pseudo-time
(time {\color{black} multiplied with} a unit diffusion coefficient), which should be distinguished from the physical time
$t$ in the CFD--DEM formulation.  To solve the problem above, we recall that the fundamental
solution (also called Green's function) of the diffusion equation in free space is~\citep[e.g.,
][Chapter 11]{haberman12app}:
\begin{equation}
 G(\mathbf{x}, \tau) = \frac{1}{ (4 \pi \tau)^{3/2}} \exp \left[ -\frac{\mathbf{x}^T \mathbf{x}}{4 \tau} \right] ,
\end{equation}
which is basically the solution to Eq.~(\ref{eq:diffusion}) with the initial condition
\(\varepsilon(\mathbf{x}, 0) = \delta(\mathbf{x})\), with \(\delta(\mathbf{x})\) being Dirac delta
function. Based on the Green's function, the solution to Eq.~(\ref{eq:diffusion}) with initial
conditions in Eq.~(\ref{eq:diffusion-b}) is:
\begin{equation}
 \varepsilon(\mathbf{x}, \tau) = \int_{\mathbb{R}^3} G(\mathbf{x}-\boldsymbol{\xi}, \tau) \; \varepsilon_0 (\boldsymbol{\xi}) d \boldsymbol{\xi}.
 \label{eq:diff-infty}
\end{equation}

\begin{figure}[!htbp]
 \centering \includegraphics[width=0.8\textwidth]{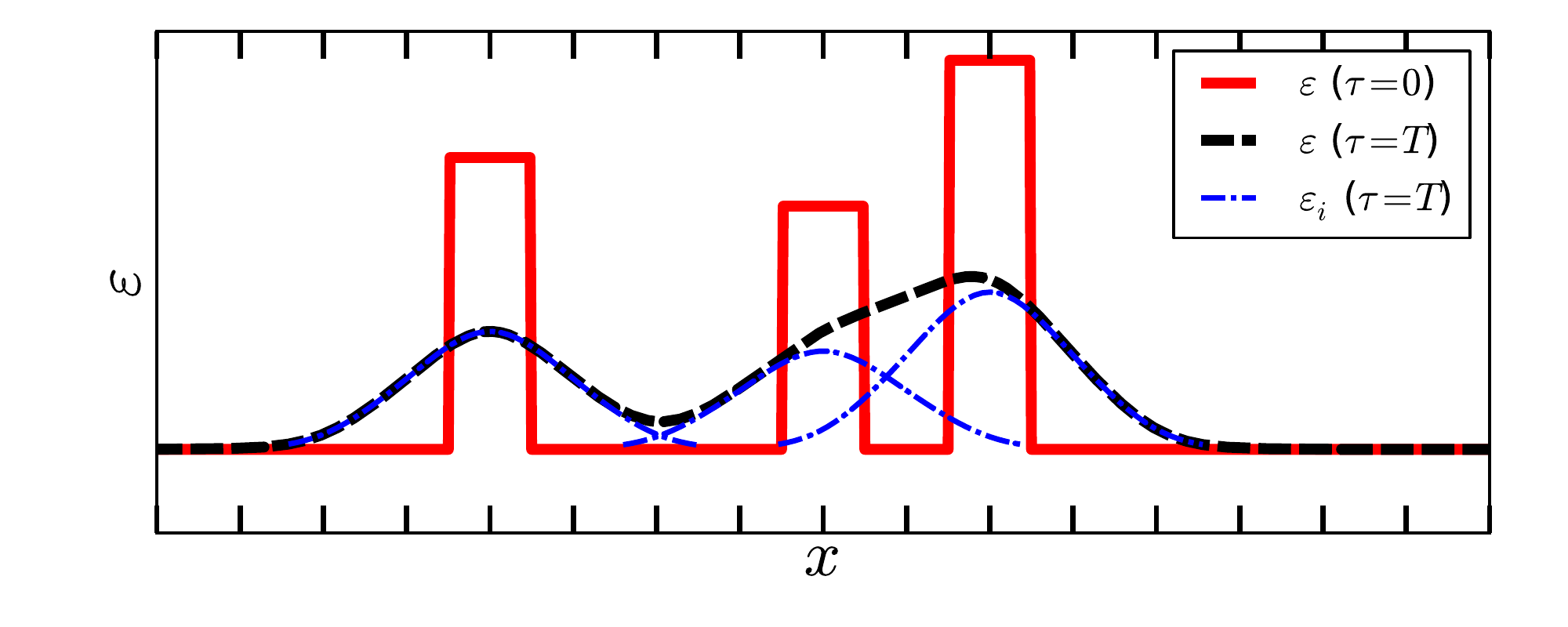}
 \caption{ 
  \label{fig:diffNp}
  The solution to the diffusion equation for the initial condition of a linear combination of
  {\color{black} shifted top-hat functions, which are approximate representations of delta
    functions on a mesh with the same integral}. The initial condition $\varepsilon(\tau=0)$ is obtained from
  the procedure of the PCM; $\varepsilon_i(\tau=T)$ are the diffused solutions corresponding to each
  shifted top-hat functions; \(\varepsilon(T)\) is the superposition of \(N_p\) Green's functions,
  which is the result obtained with the proposed averaging procedure.}
\end{figure}

Consider the initial condition consisting of a linear combination of \(N_p\) shifted delta functions
centered at \(\mathbf{x}_i\), where \(i = 1, \cdots, N_p\),  that is,
\begin{equation}
 \varepsilon_0(\mathbf{x}) = \sum_{i=1}^{N_p} V_{p, i} \; \delta(\mathbf{x} - \mathbf{x}_i),
 \label{eq:c-delta}
\end{equation}
where \(V_{p, i}\) are multiplier coefficients for each shifted delta function. The interpretation
of \(V_{p, i}\) as particle volume and \(N_p\) as number of particles will be evident shortly.
Plugging in  Eq.~(\ref{eq:c-delta}) to Eq.~(\ref{eq:diff-infty}) yields the solution to the
diffusion equation~(\ref{eq:diffusion}) with the initial condition of superimposed shifted delta
functions:
\begin{align}
 \varepsilon(\mathbf{x}, \tau) \equiv \sum_{i=1}^{N_p} \varepsilon_i = \sum_{i=1}^{N_p} V_{p, i} \; G(\mathbf{x} - \mathbf{x}_i, \tau) ,
\end{align}
which is a linear combination of the $N_p$ Green functions. This is illustrated in
Fig.~\ref{fig:diffNp}.

Consider the solution at a fixed pesudo-time \(\tau = T\), denoted as
\(\hat{\varepsilon}(\mathbf{x}) \equiv \varepsilon(\mathbf{x}, T)\). The solution is rewritten as:
\begin{equation}
 \hat{\varepsilon}(\mathbf{x}) = \sum_{i=1}^{N_p} V_{p,i} \; G_i \; ,
 \label{eq:solutionDiff}
\end{equation}
where
\begin{equation}
 G_i = G(\mathbf{x} - \mathbf{x}_i) = \frac{1}{(4 \pi T)^{3/2}} \exp \left[ -\frac{(\mathbf{x}-\mathbf{x}_i)^T (\mathbf{x}-\mathbf{x}_i)}{4T} \right] .
 \label{eq:solutionGi}
\end{equation}
Comparing Eqs.~(\ref{eq:solutionDiff}) and (\ref{eq:solutionGi}) with Eqs.~(\ref{eq:superpo}) and
(\ref{eq:hi}), the equivalence between diffusion based and the statistical kernel based coarse
graining is established with \(b = \sqrt{4T}\). The physical interpretation is that given \(N_p\)
particles centered at \(\mathbf{x}_i\)  with volume \(V_{p, i}\), the following two ways of
calculating solid volume fraction field are equivalent:
\begin{enumerate}
\item
compute the solid volume fraction field \(\varepsilon_0\) by using the PCM, and solve
Eq.~(\ref{eq:diffusion}) until \(\tau = T\) with \(\varepsilon_0\) as initial condition; and
\item
  use the Gaussian kernel based averaging according to Eqs.~(\ref{eq:solutionDiff}) and
  (\ref{eq:solutionGi}) with bandwidth $b = \sqrt{4T}$.
\end{enumerate}

\subsubsection{Treatment of Physical Boundaries}
The equivalence shown above holds for interior domains where boundary effects are not present. For
particles near boundaries, we consider the same diffusion equation as Eq.~(\ref{eq:diffusion}) on a
semi-infinite domain \(\mathcal{D} = \{(\mathbf{x}, \tau) | P(\mathbf{x}) \ge 0, \tau>0 \}\), where
\(P(\mathbf{x}) = 0\) is the equation of the boundary plane. Zero-gradient Neumann boundary
condition, i.e., \( \partial \varepsilon / \partial n = 0\), is specified at the boundary, where $n$
is the normal direction of the plane.

The solution to Eq.~(\ref{eq:diffusion}) at time \(\tau = T\) with initial condition
\begin{equation*}
 \varepsilon_0 = \sum_{i=1}^{N_p} V_{p, i} \delta(\mathbf{x}-\mathbf{x}_i)
\end{equation*}
 can be obtained by constructing a Green's function that satisfy the zero-gradient condition on the
 boundary plane. It turns out that such a Green's function, denoted as \( \tilde{G}_i \), can be
 found by the method of images~\citep{haberman12app}:
\begin{align}
  \tilde{G}_i &
  = G(\mathbf{x}-\mathbf{x}_i, T) + G(\mathbf{x}-\mathbf{x}_i', T) \\
  & = \frac{1}{ (4 \pi T)^{3/2}} \left( \exp \left[ -\frac{(\mathbf{x}-\mathbf{x}_i)^T (\mathbf{x} -
        \mathbf{x}_i)}{4T} \right] + \exp \left[ -\frac{(\mathbf{x}-\mathbf{x}_i')^T
        (\mathbf{x}-\mathbf{x}_i')}{4T} \right] \right),
 \label{eq:diffHalf}
\end{align}
where \(\mathbf{x}'_i\) is the image of \(\mathbf{x}_i\) with respect to the boundary plane
\(P(\mathbf{x}) = 0\). The solution to the diffusion equation on a domain bounded by plane \(
P(\xmb{x}) = 0 \) is:
\begin{equation}
 \hat{\varepsilon}(x) = \sum_{i=1}^{N_p} V_{p,i} \; \tilde{G}_i \; .
 \label{eq:solution-bound}
\end{equation}
Clearly, the solution \(\tilde{G}_i\) is equivalent to the kernel function \(\tilde{h}_i\) in
Eq.~(\ref{eq:hi-bound}) when \(b=\sqrt{4T}\).

The equivalence can be extended to the general case where the domain is bounded in all three
directions and a particle is located near the corner, i.e., it is close to many, possibly
non-orthogonal boundary planes (see Fig.~\ref{fig:wall-ext}). On such a domain, with no-flux
boundary conditions the Green's function for the diffusion equation is a summation of infinite
series~\citep{haberman12app}:
\begin{equation}
 \tilde{G}_i 
 = \sum_{n = 0}^{\infty} G^{(n)}(\mathbf{x}-\mathbf{x}^{(n)}_i, T) ,
 \label{eq:non-orth}
\end{equation}
which corresponds exactly with the summation of infinite series of kernel functions in
Eq.~(\ref{eq:diff-infty}). This is not surprising since this solution is obtained with the method of
images, which is the same idea based on which the kernel function in Eq.(\ref{eq:diff-infty}) is
obtained.  In summary, the diffusion-based averaging procedure as described in
Section~\ref{sec:diff-alg} is theoretically equivalent to the statistical kernel method presented in
Section~\ref{sec:kernel}.

Numerically the equivalence between the two methods holds up to the mesh discretization accuracy.
If the mesh used to obtain \(\varepsilon_0\) is sufficiently small, or equivalently, if the initial
condition obtained via PCM used for the diffusion equation is approximately a linear combination of
shifted delta functions corresponding to individual particles, then the two methods are numerically
equal. We emphasize that the diffusion-based method can handle both particles in the interior domain
and those near the boundaries in a \emph{unified} manner, i.e., by solving a diffusion equation with
no-flux boundary conditions. The equivalence to statistical kernel function is valid for domains of
arbitrary shapes, for both interior particles and particles close to arbitrary boundaries. This
characteristics of the current method distinguishes itself from its theoretically equivalent
counterpart, the statistical kernel-based averaging, where obtaining the Green's functions for
particles located near corners in a generic three-dimensional domain can be tedious.

\section{A Priori Numerical Tests}
\label{sec:apriori}

In this section, the diffusion-based averaging algorithm described in Section~\ref{sec:cg} is
used to obtain coarse grained solid volume fraction $\varepsilon_s$ fields given a quasi-random
distribution of particles in the computational domain. Note that even though we occasionally refer
to ``CFD mesh'' in the text, no CFD or DEM simulations are performed in these tests, and the
particles do not move.

In Section~\ref{sec:apriori-eq}, we first compare the coarse grained $\varepsilon_s$ fields obtained
by using the diffusion-based method and those obtained with Gaussian kernel functions to verify the
theoretical equivalence of the two shown in Section~\ref{sec:connect}. In some CFD--DEM simulations,
the flow field resolution requirements may necessitate using cells that are comparable to or smaller
than the particle diameters in certain regions. As a consequence, using CFD mesh for averaging
as is done in PCM and DVPM can lead to two potential problems: unphysically large $\varepsilon_s$ in
small cells and mesh-dependent $\varepsilon_s$ fields.  To demonstrate the merits of the proposed
method, tests in Sections~\ref{sec:apriori-comp}--\ref{sec:apriori-independence} aim to show its
capabilities to produce smooth coarse grained fields on meshes with small cells, to handle stretched
and unstructured meshes, and to produce mesh-independent results. Comparisons are made with other
averaging methods when applicable.

\begin{figure}[!htpb]
  \centering
  \subfloat[][interior particles]{
    \includegraphics[width=0.45\textwidth]{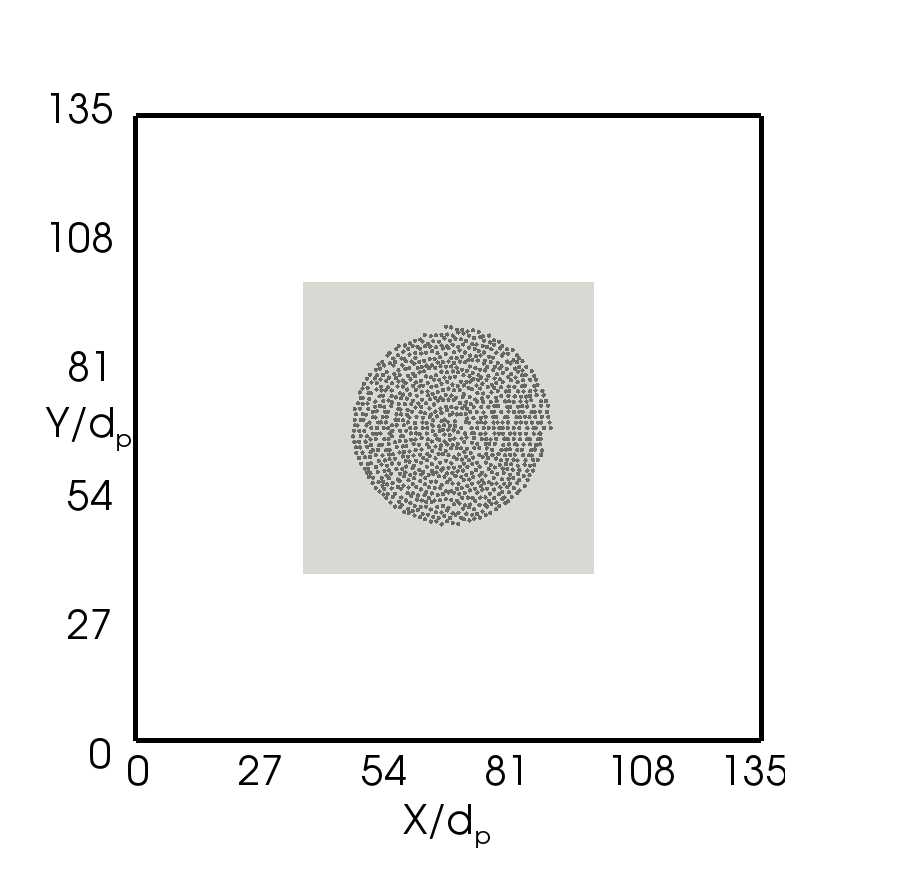}
  }
  \hspace{0.01\textwidth}
  \subfloat[][near-boundary particles]{
    \includegraphics[width=0.45\textwidth]{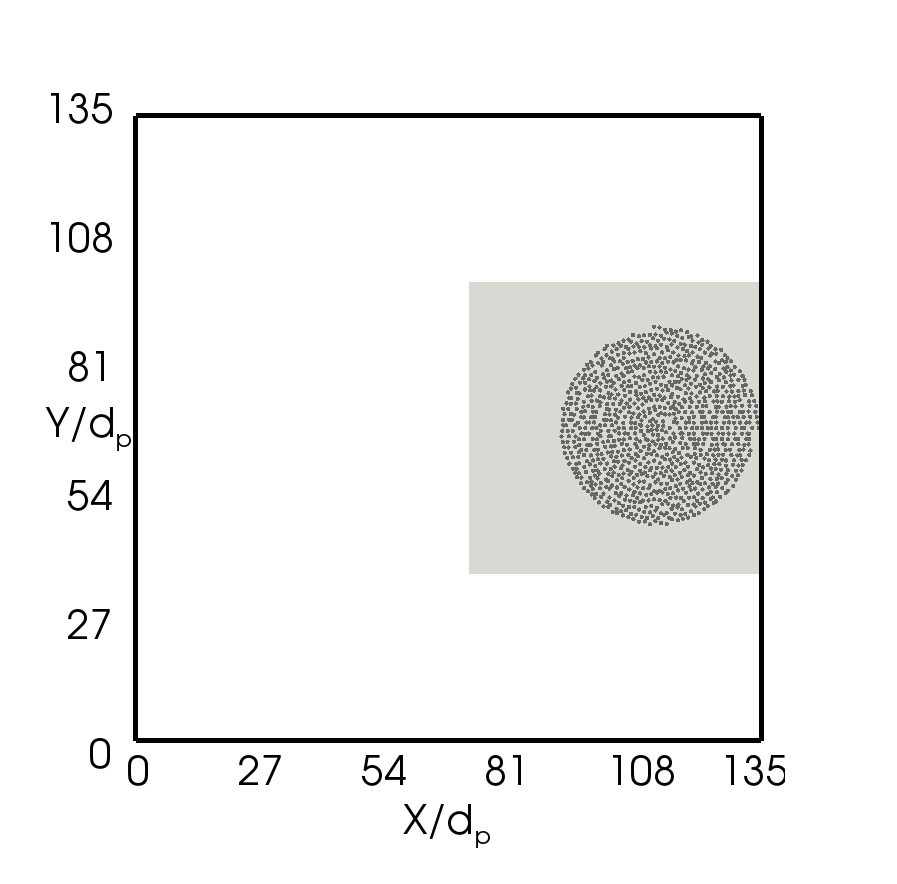}
  }
  \caption{Distribution of particles in computational domain showing particles located in (a)
    interior cells and (b) near-boundary cells. The size of the computational domain is $L_x \times
    L_y = 135d_p \times 135d_p$ (dimensions in $x$ and $y$ directions, respectively). Shades
    indicate regions of interests, coarse graining fields for which will be presented in subsequent
    plots. Thick lines indicated the boundaries of the computational domain.}
  \label{fig:apriori-domain}
\end{figure}

All simulations presented below are performed with 1000 spherical particles with quasi-random
spatial distribution in the computational domain.  In the verification tests presented in
Section~\ref{sec:apriori-eq}, two representative distributions of particles are used to evaluate the
performance of the proposed method for particles located in interior cells and those located near
boundaries.  The two configurations are displayed in Figs.~\ref{fig:apriori-domain}(a)
and~\ref{fig:apriori-domain}(b), respectively. In both cases the sizes of the domain are $L_x \times
L_y = 135d_p \times 135d_p$, where $L_x$ and $L_y$ are the dimensions in $x$ and $y$ directions,
respectively. Simulations are performed on meshes with only one layer of cells in $z$-direction. The
thickness of the cells is the same as the particle diameter, and all particles are located on the
same plane normal to the $z$-axis.  No-flux conditions are specified at all boundaries to ensure
conservation of particle mass in the coarse graining procedure.

\subsection{Verification of Equivalence to Gaussian Kernel Averaging}
\label{sec:apriori-eq}

In this test, solid volume fraction field is obtained from the particle distribution by using both
statistical kernel averaging and the diffusion-based method to demonstrate the equivalence
between the two. The mesh resolution is $N_x \times N_y = 45 \times 45$, where $N_x$ and $N_y$ are
the numbers of cells in $x$ and $y$ directions, respectively. The bandwidth $b$ used for the
statistical kernel method is $6d_p$, and the corresponding diffusion time $T$ is $9d_p^2$ for the
diffusion-based method according to $b=\sqrt{4T}$. This test is performed for both interior
particles and near-boundary particles.

\begin{figure}[!htpb]
  \centering
  \subfloat[][mesh and particle distribution]{
    \includegraphics[width=0.42\textwidth]{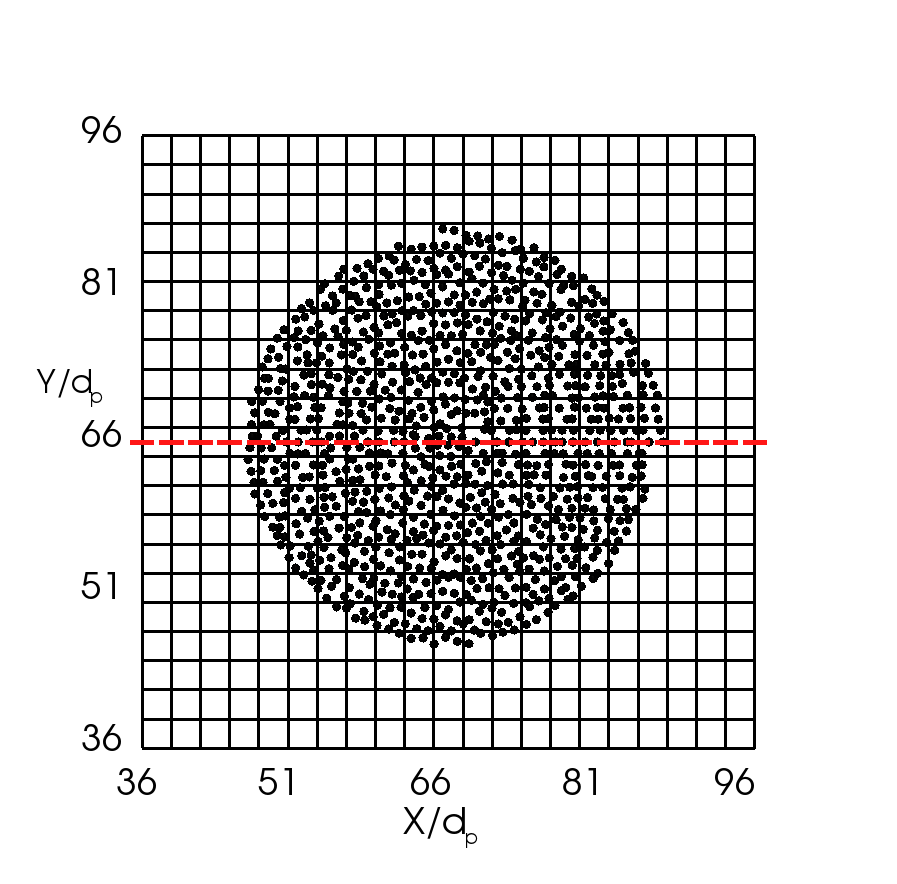}
  }
  \hspace{0.01\textwidth}
  \subfloat[][solid volume fraction]{
    \includegraphics[width=0.45\textwidth]{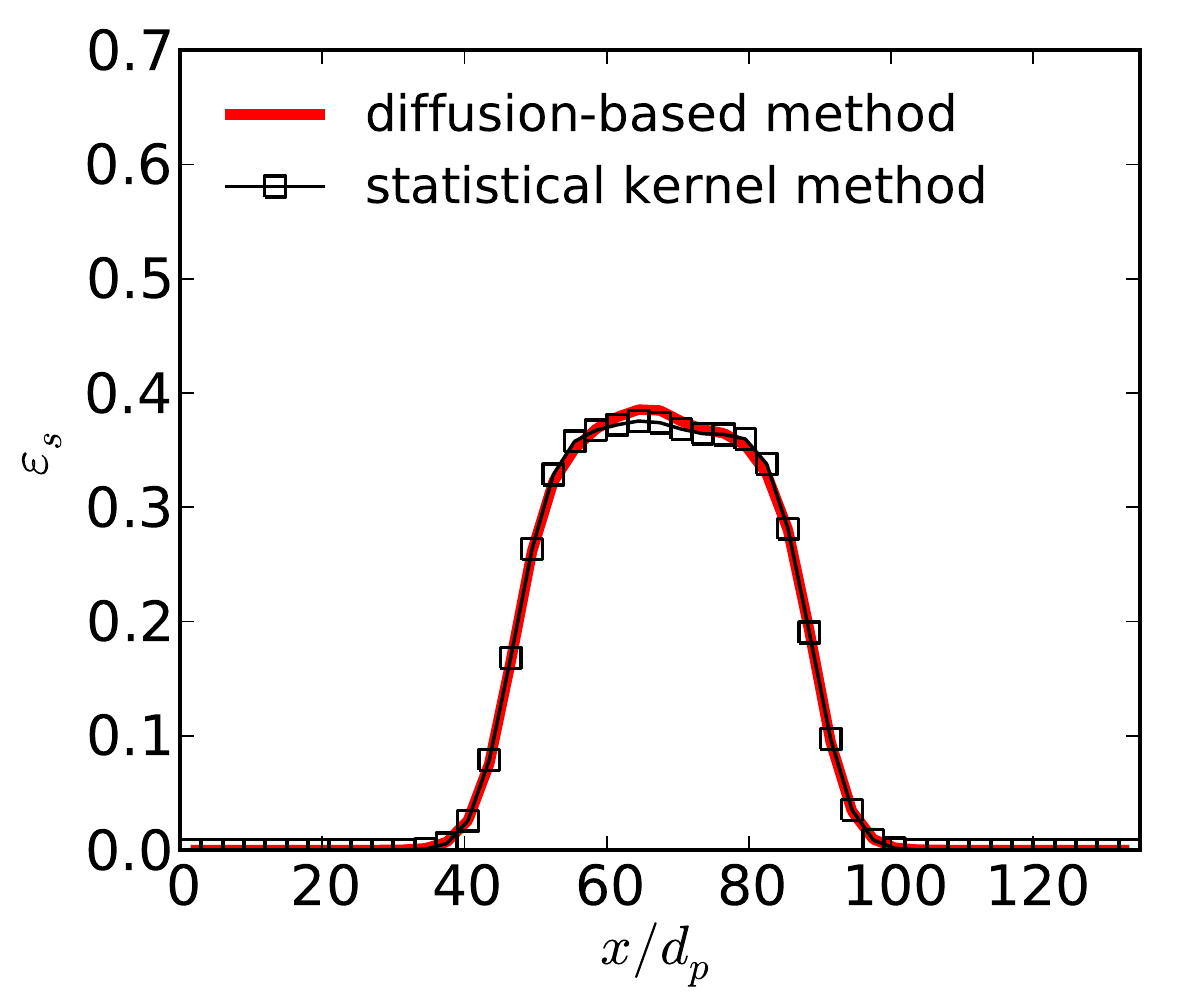}
  }
  \caption{Equivalence between statistical kernel method and diffusion-based method for interior
    particles, showing (a) the particle distribution and the mesh used for the averaging, and
    (b) the solid volume fraction along the dashed line indicated in panel (a). The same bandwidth
    $b=6d_p$ is used for both methods. Due to the randomness of the particle
      distribution, exact symmetry of the obtained volume fraction with respect to the centerline is
      not guaranteed. This is not a deficiency of the coarse-graining method. The same comment applies
      to Figs.~\ref{fig:diff-PCM}--\ref{fig:diff-dt-study}.}
  \label{fig:apriori-eq1}
\end{figure}

\begin{figure}[!htpb]
  \centering
  \subfloat[][mesh and particle distribution]{
    \includegraphics[width=0.42\textwidth]{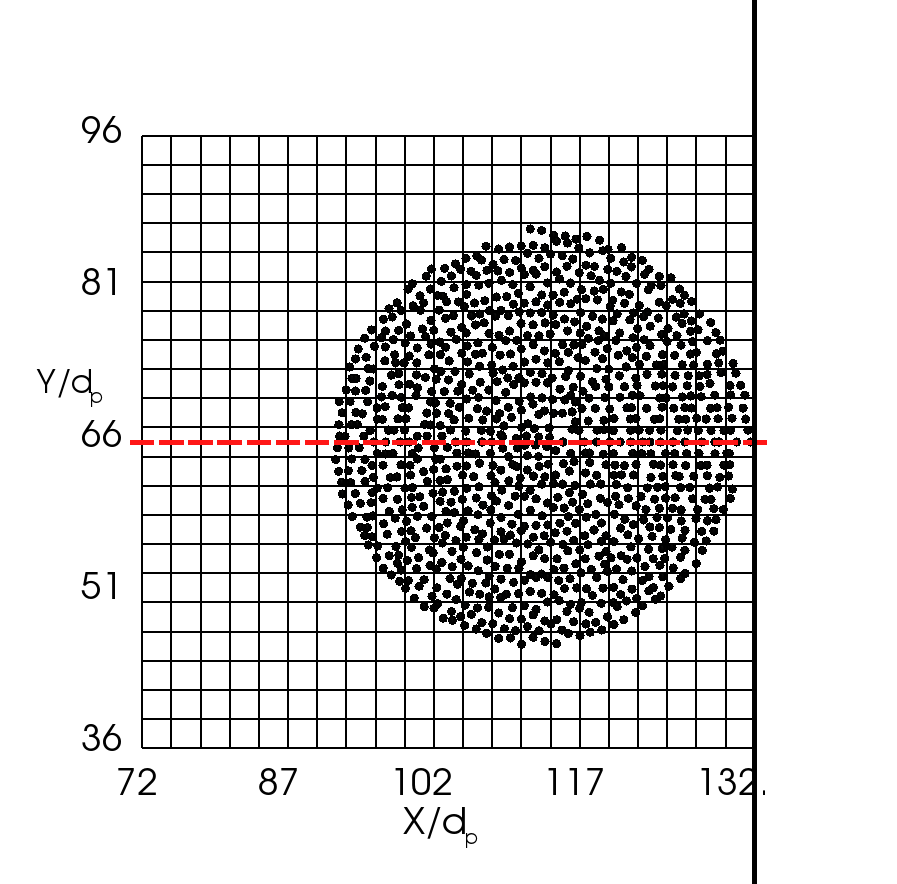}
  }
  \hspace{0.01\textwidth}
  \subfloat[][solid volume fraction]{
    \includegraphics[width=0.45\textwidth]{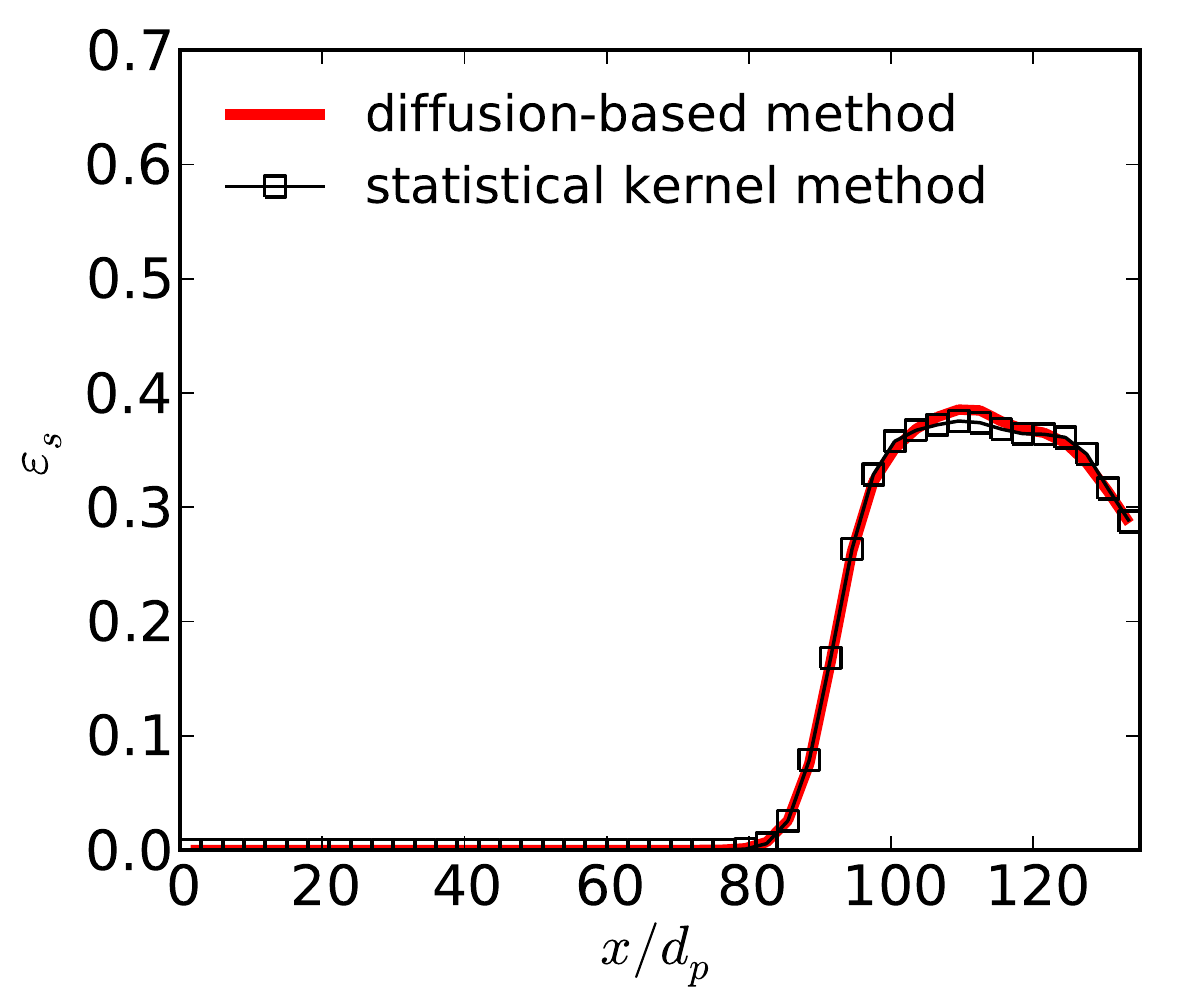}
  }
  \caption{Equivalence between statistical kernel method and diffusion-based method for
    near-boundary particles, showing (a) the particle distribution and the mesh used for the coarse
    graining, and (b) the solid volume fraction along the dashed line indicated in Panel~(a).}
  \label{fig:apriori-eq2}
\end{figure}

For interior particles, the solid volume fraction fields along the centerline of the domain
(indicated by the dashed line in Fig.~\ref{fig:apriori-eq1}(a)) obtained by using both methods are
displayed in Fig.~\ref{fig:apriori-eq1}.  The domain of interest (shaded region in
Fig.~\ref{fig:apriori-domain}(a)) is shown in Fig.~\ref{fig:apriori-eq1}(a), and the coarse grained
$\varepsilon_s$ fields are presented in Fig.~\ref{fig:apriori-eq1}(b). It can be seen that the
$\varepsilon_s$ fields obtained by using statistical kernel-based averaging and that by using
the diffusion-based method agree very well. For the configuration with particles located near
boundaries as shown in Fig.~\ref{fig:apriori-domain}(b), the mesh and detailed particle distribution
are shown in Fig.~\ref{fig:apriori-eq2}(a), and Fig.~\ref{fig:apriori-eq2}(b) displays the
$\varepsilon_s$ fields obtained by using both methods. Again, that the coarse grained
$\varepsilon_s$ fields in the simulation using both methods are consistent.

From the results above, it can be seen that the statistical kernel method with Gaussian kernels and
the proposed method are indeed equivalent for both interior and near-boundary particles. Note that
the boundary effect is accounted for in the Gaussian kernel averaging by using the image
kernels in Eq.~(\ref{eq:hi-bound}).  While the kernal function in the statistical kernel method
needs to be modified to accommodate the proximity of particles to the boundary (see
Section~\ref{sec:kernel-withbnd}), the diffusion-based method remains the same for both the interior
and the near-boundary particles by using no-flux boundary conditions.  This simplicity in formulation
and implementation is an important improvement of the proposed method over statistical kernel-based
averaging method.

\subsection{Comparison with Existing Coarse Graining Methods}
\label{sec:apriori-comp}

To demonstrate the advantages of the diffusion-based coarse graining method, the comparison of the
diffusion-based method with PCM, DPVM, and two-grid formulation is performed. The set-up and
parameters, including the computational domain, mesh resolution, particle distribution, bandwidth,
diffusion time are the same as in Section~\ref{sec:apriori-eq}.  Since the performances of these
coarse graining methods are similar for both interior and near-boundary particles, only the results
for interior particles are presented below.

Figure~\ref{fig:diff-PCM} shows the comparison of the $\varepsilon_s$ fields obtained by using PCM,
DPVM, and the diffusion-based method. It can be seen that the largest $\varepsilon_s$ in the
$\varepsilon_s$ field computed by PCM is 0.64, which is clearly unphysical as it exceeds the maximum
possible value (0.606) for close-packed spherical particles in a two-dimensional
domain~\citep{chang10si}.  This is attributed to the inaccuracy of the PCM when the centroid of a
particle is located near the boundaries of a cell as discussed in Section~\ref{sec:pcm}. The maximum
$\varepsilon_s$ value in the DPVM results is smaller than that of PCM, since the volume of a
particle is divided among all cells it intersects with. Although the $\varepsilon_s$ field obtained
by using DPVM is smoother than the PCM results, the overall fluctuations and gradients are still
large. Considering the statistically uniform particle distribution in the domain, which is evident
from visual observations of Fig.~\ref{fig:apriori-eq1}(a), these fluctuations cannot be justified
physically, and thus are considered artifacts introduced by the averaging procedure. In
contrast, the $\varepsilon_s$ field obtained by using the diffusion-based method is much smoother
with a flat region in the middle (between $x/d_p$ = 50 and 80), confirming the visual observation of
statistically uniform particle distribution near the core of the region occupied by the particles.

\begin{figure}[!htpb]
  \centering
  \includegraphics[width=0.5\textwidth]{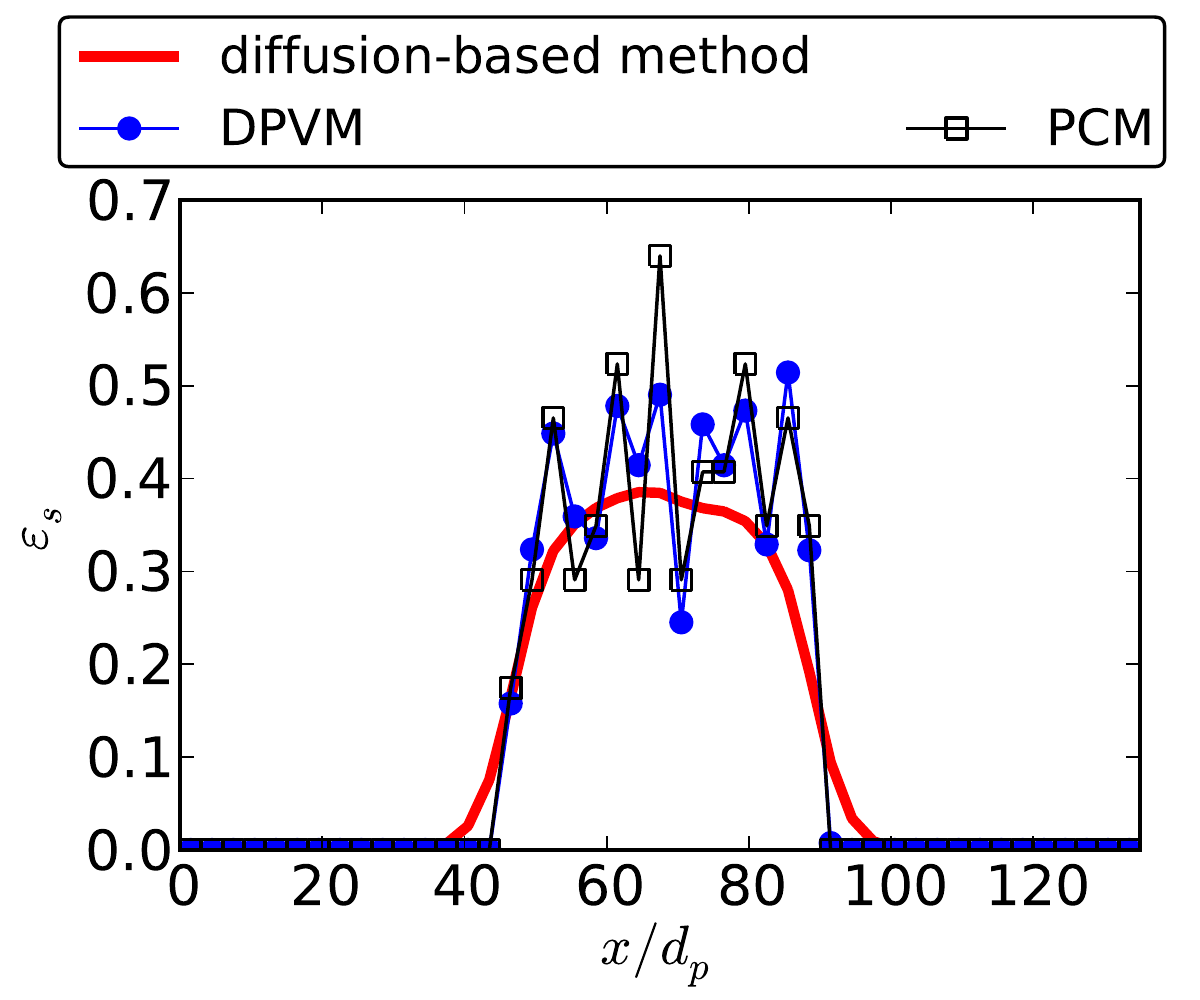}
  \caption{Comparison of coarse-grained solid volume fraction obtained by using PCM, DPVM, and
    diffusion-based averaging method. Shown is the solid volume fraction along the horizontal
    centerline of the domain indicated by the dashed line in
    Fig.~\ref{fig:apriori-eq1}(a).\label{fig:diff-PCM} }
  \end{figure}

\begin{figure}[!htpb]
  \centering
  \subfloat[][mesh and particle distribution]{
    \includegraphics[width=0.45\textwidth]{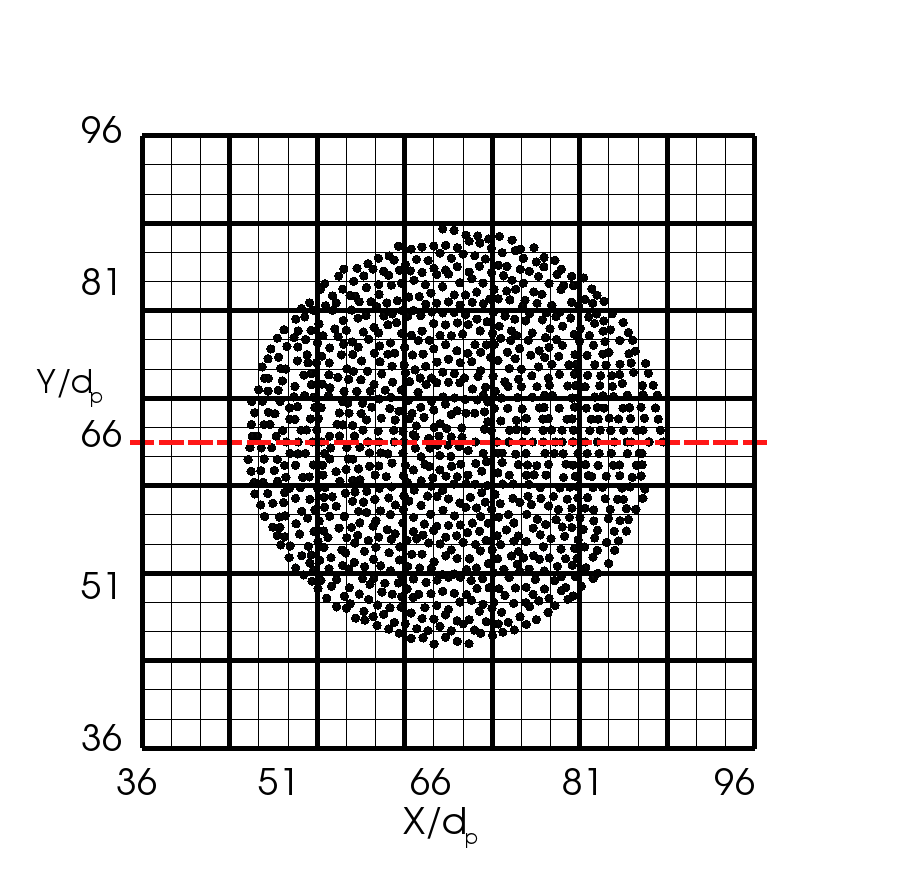}
  }
  \hspace{0.01\textwidth}
  \subfloat[][solid volume fraction]{
    \includegraphics[width=0.45\textwidth]{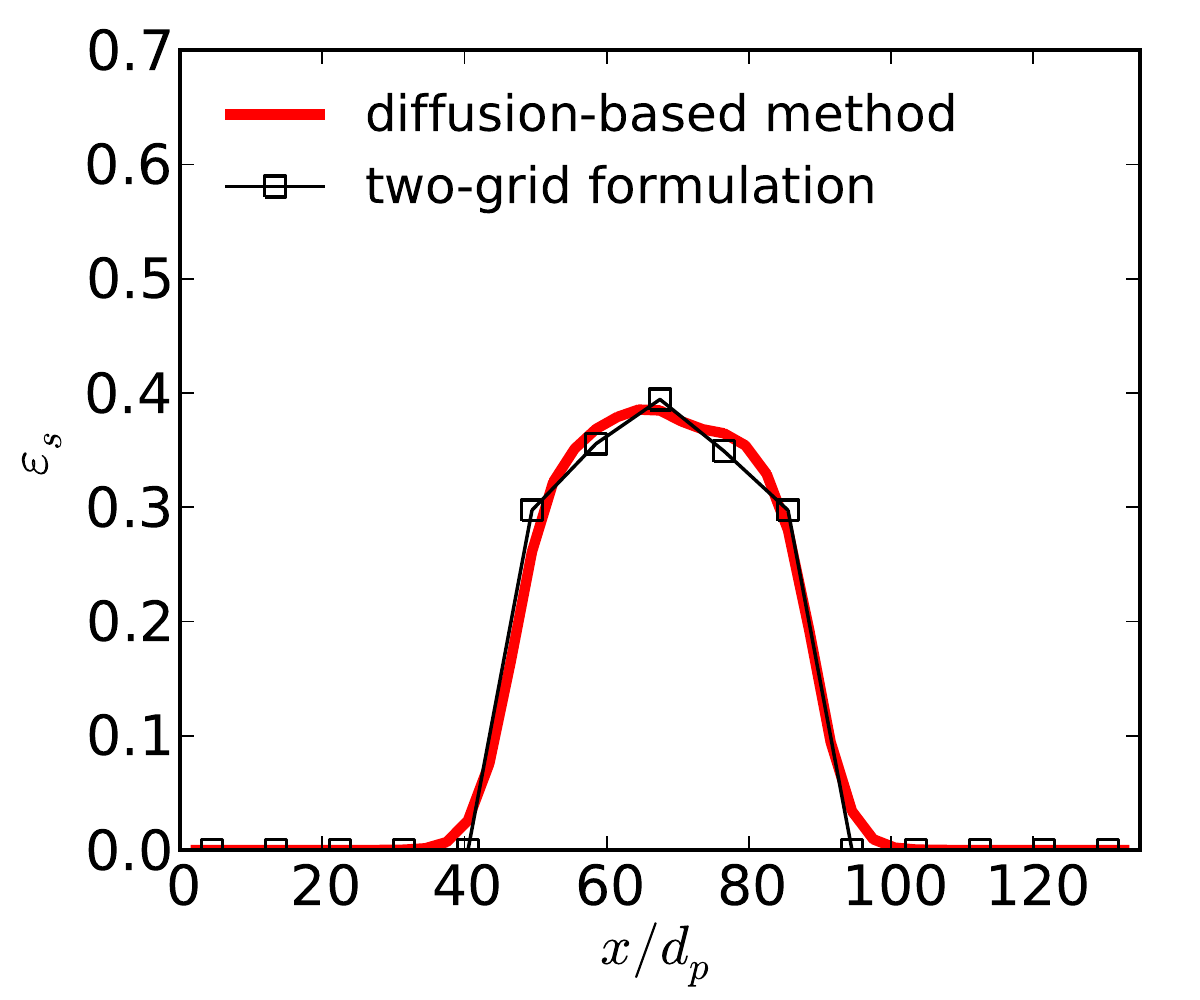}
  }
  \caption{
    \label{fig:diff-tg}
    Comparison of the diffusion-based averaging method and the two-grid formulation. Panel
    (a) displays the mesh and the particle distribution in the shaded region of
    Fig.~\ref{fig:apriori-domain}(a). Each cell in the coarse graining mesh, indicated in thick
    lines, contains exactly $3 \times 3$ CFD cells. Although CFD simulations are not performed here,
    the mesh is used for the averaging with the diffusion-based method. Panel (b) shows the
    solid volume fraction along the horizontal centerline of the domain (dashed line in panel~(a)).}
\end{figure}

  The comparison of diffusion-based method and two-grid formulation is presented in
  Fig.~\ref{fig:diff-tg}. The configuration of the coarse graining mesh is displayed in
  Fig.~\ref{fig:diff-tg}(a), showing that each cell in the coarse graining mesh contains exactly $3
  \times 3$ CFD cells.  From the $\varepsilon_s$ profiles along the horizontal centerline presented
  in Fig.~\ref{fig:diff-tg}(b), both methods are able to produce smooth solid volume fraction fields
  that are consistent with each other. Noted that the resolution of the coarse graining mesh for
  two-grid formulation is lower than that of the CFD mesh (with much larger cells), which explains
  why the results obtained by with the diffusion-based method is smoother.  The $\varepsilon_s$
  values on the CFD mesh is not directly obtained, and interpolations~(also called mapping) from the
  coarse-graining mesh to the CFD mesh are needed to obtain $\varepsilon_s$ in CFD
  cells~\citep{Deb13-ano}. Linear interpolation is used to obtain the results presented in
  Fig.~\ref{fig:diff-tg}(b).  In contrast, in the diffusion-based method the coarse grained
  $\varepsilon_s$ field is obtained directly on the fine mesh without the need of interpolations
  between two meshes. Another problem in the implementation of the two-grid formulation is the
  agglomeration of the unstructured mesh, which has been discussed in Section~\ref{sec:two-grid}.
  Therefore, despite the fact that the two methods give consistent coarse grained $\varepsilon_s$
  fields in this case, we argue that the diffusion-based method is preferred when used in general
  CFD solvers due to its straightforward implementation on arbitrary meshes.

\begin{figure}[!htpb]
  \centering
  \subfloat[][PCM]{
    \includegraphics[width=0.4\textwidth]{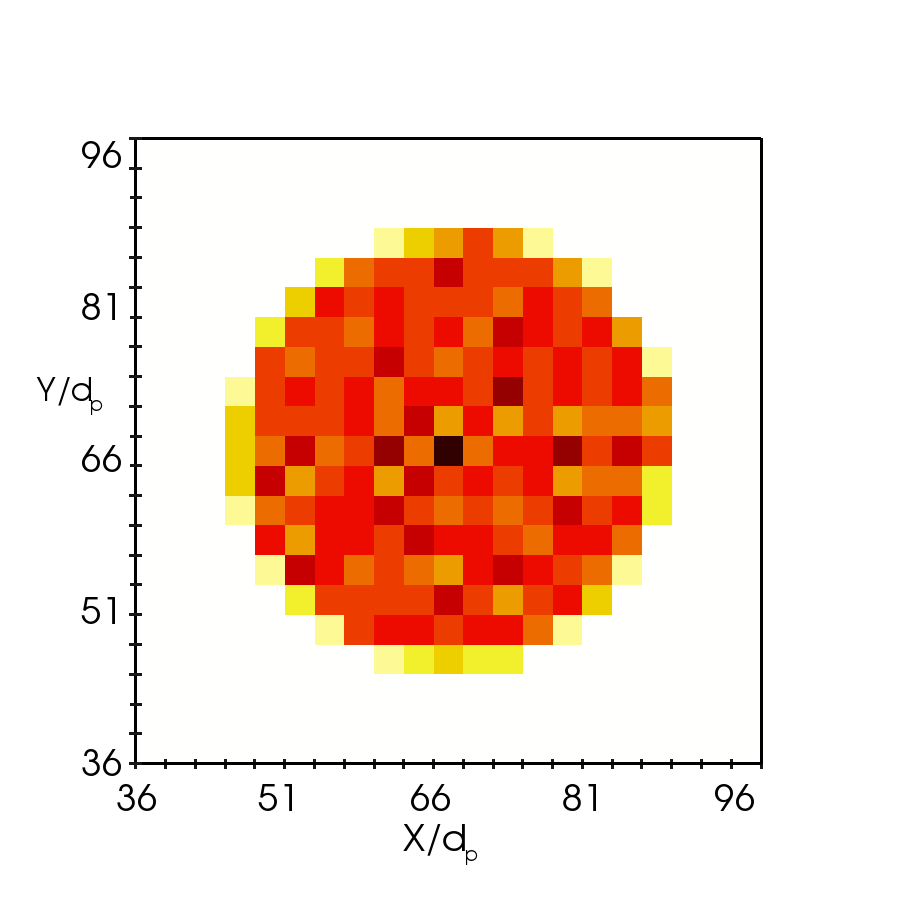}
  }
  \hspace{0.01\textwidth}
  \subfloat[][DPVM]{
    \includegraphics[width=0.4\textwidth]{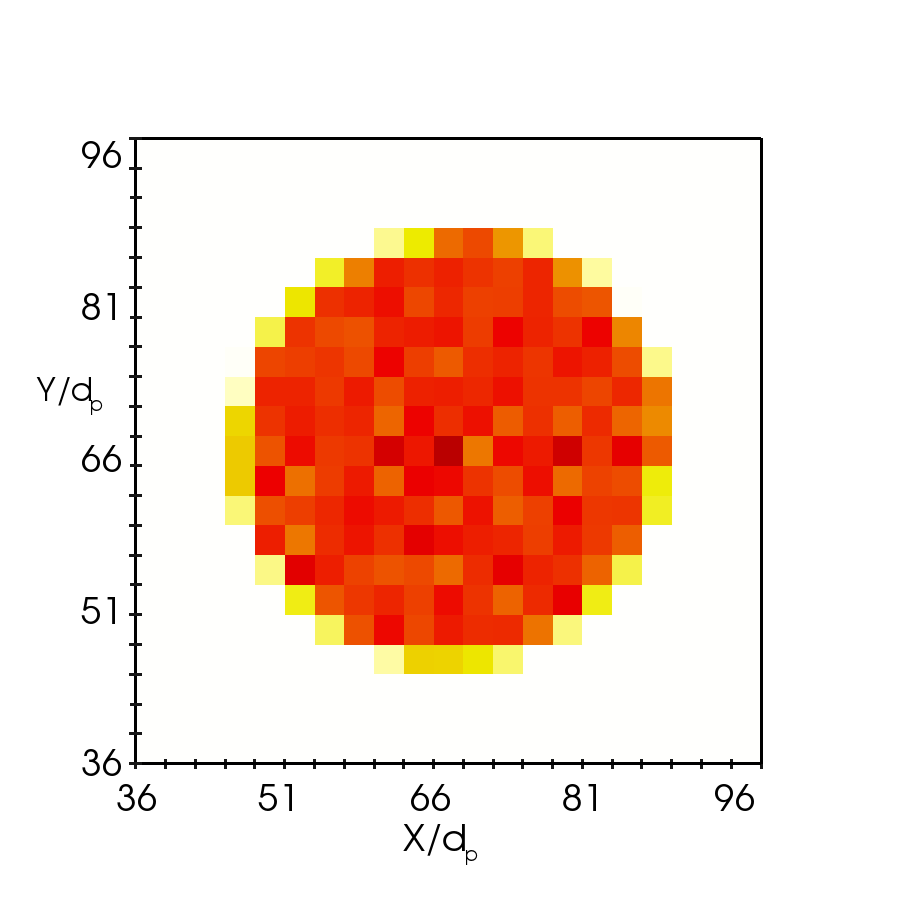}
  }
  \vspace{0.01\textwidth}
  \subfloat[][two-grid formulation]{
    \includegraphics[width=0.4\textwidth]{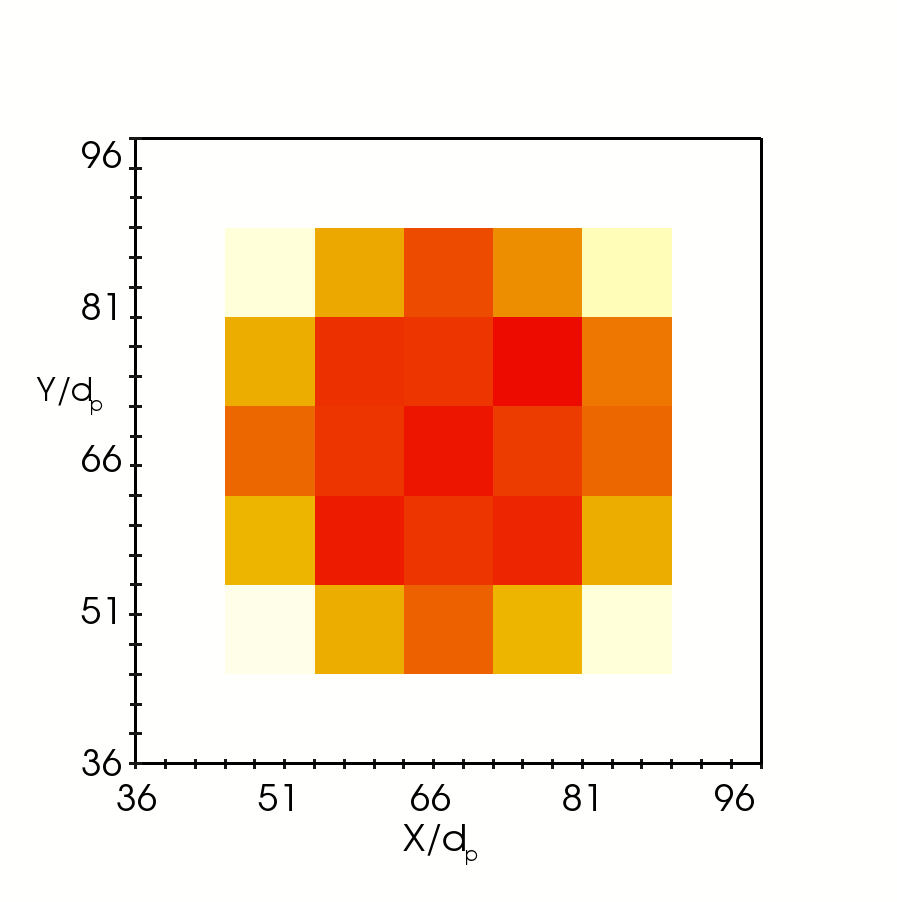}
  }
  \hspace{0.01\textwidth}
  \subfloat[][diffusion-based method]{
    \includegraphics[width=0.4\textwidth]{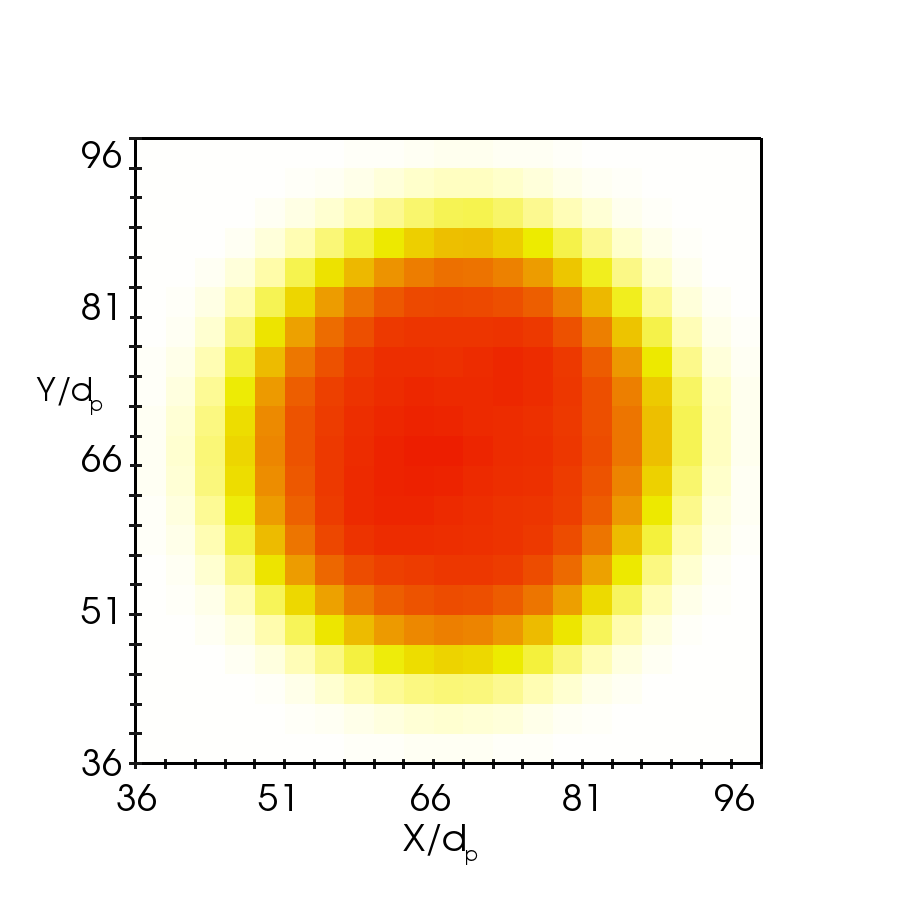}
  }
  \vspace{0.01\textwidth}
  \subfloat[][]{
    \includegraphics[width=0.6\textwidth]{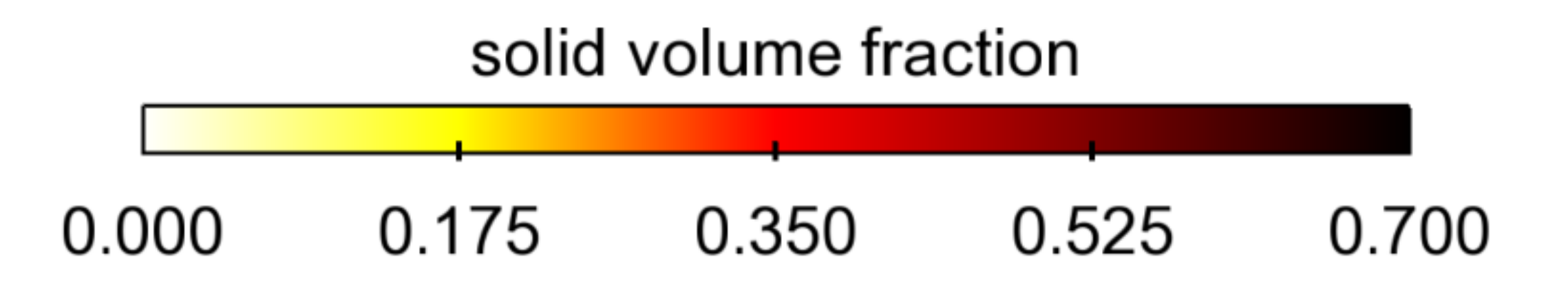}
  }
  \caption{
    \label{fig:diff-contour}
    Comparison of the solid volume fraction $\varepsilon_s$ fields obtained by using the four coarse
    graining method: (a) PCM, (b) DPVM, (c) the two-grid formulation, and (d) the diffusion-based
    method. The particle distribution and the meshes used to obtain the plots for PCM, DPVM, and the
    diffusion-based method are shown in Fig.~\ref{fig:apriori-eq1}(a) and those for the two-grid
    formulation are shown in Fig.~\ref{fig:diff-tg}(a).}
\end{figure}

  To summarize the comparison, solid volume fraction contours in the two-dimensional domain obtained
  by using the four methods, PCM, DPVM, two-grid formulation, and diffusion-based method, are
  presented in Fig.~\ref{fig:diff-contour}.  In these plots, each cell is colored according to the
  corresponding $\varepsilon_s$ value in the cell, and no interpolations are used to obtain the
  contours. Consequently, the meshes used in each case (presented above) are clearly distinguishable
  from the plots.  As is evident from the contours, extreme high or low values (and thus large
  gradients) are most frequent in the PCM results, and also occasionally occur in the DPVM
  result. In contrast, no such extreme values are present in the plot obtained with the two-grid
  formulation and that from the diffusion-based method. In the two-grid formulation result, all CFD
  cells in the same coarse graining cell have the same $\varepsilon_s$ value due to the simple
  mapping scheme used (following \cite{Deb13-ano}).

\subsection{Averaging on Stretched and Unstructured Meshes}
\label{sec:apriori-meshshape}

In this subsection, the performance of the diffusion-based averaging method is evaluated on a
stretched mesh and an unstructured mesh with triangular cells. These types of meshes are commonly
used in industrial CFD simulations. Since these highly irregular meshes pose difficulties in the
averaging procedure, this is an illustration of the merits of the diffusion-based method. The
computational set-up and parameters are the same as used in Section~\ref{sec:apriori-eq}. The coarse
grained $\varepsilon_s$ field obtained by using the uniformly spaced mesh presented in
Fig.~\ref{fig:apriori-eq1}(a) is used as benchmark solution for comparison purposes.

\begin{figure}[!htpb]
  \centering
  \subfloat[][mesh and particle distribution]{
    \includegraphics[width=0.45\textwidth]{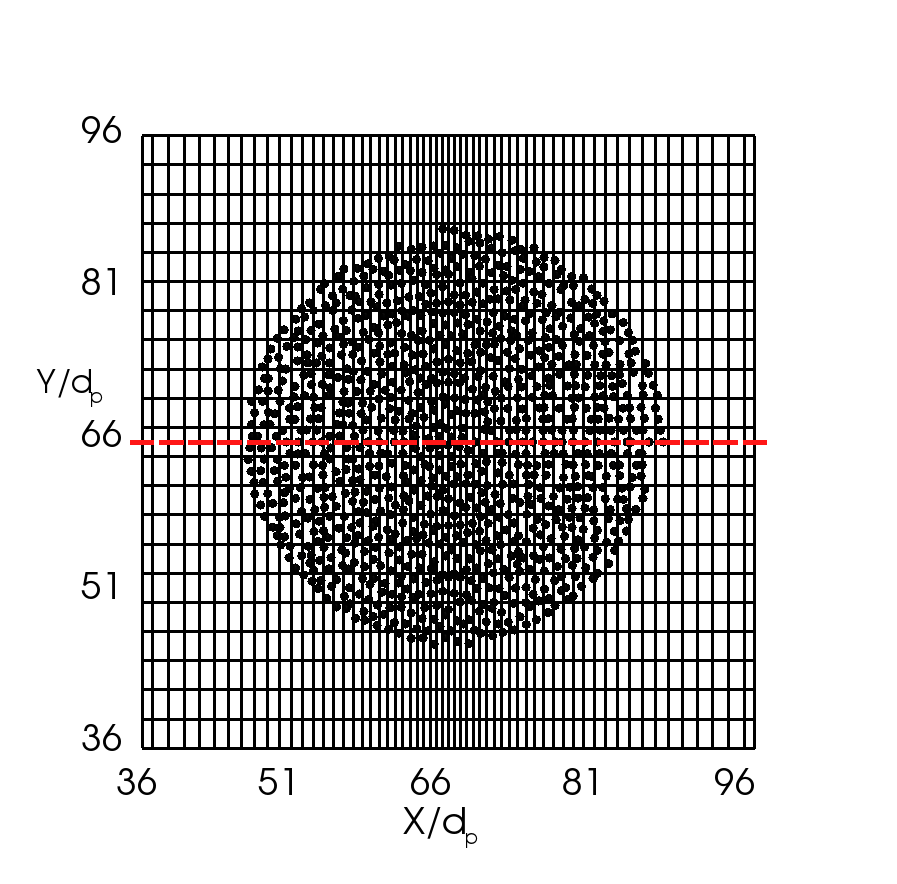}
  }
  \hspace{0.01\textwidth}
  \subfloat[][solid volume fraction]{
    \includegraphics[width=0.45\textwidth]{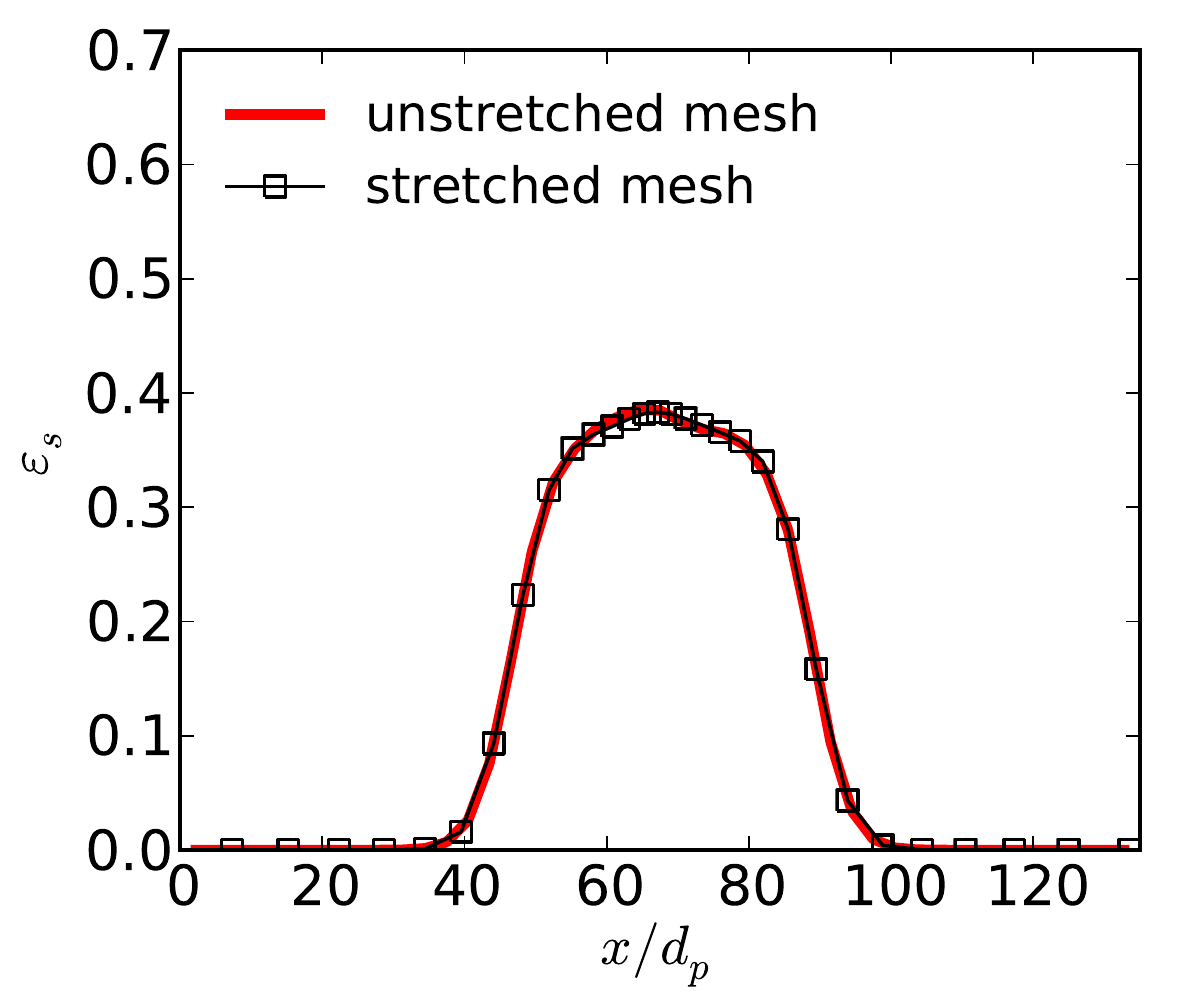}
  }
  \caption{Comparison between solid volume fraction obtained with uniform mesh and stretched mesh
    using diffusion-based averaging method. Panel (a) displays the mesh and the particle
    distribution in the shaded region in Fig.~\ref{fig:apriori-domain}(a). The mesh is stretched in
    $x$-direction with a resolution of $N_x \times N_y = 90 \times 45$. Panel (b) shows the solid
    volume fraction along the horizontal centerline of the domain (dashed line in panel~(a)).
    \label{fig:diff-str} }
\end{figure}

Figure~\ref{fig:diff-str}(a) shows the particle locations on top of the stretched mesh.  The mesh has
$N_x \times N_y = 90 \times 45$ cells with progressive refinement in $x$-direction towards the
vertical centerline, and the stretch ratio, defined as the ratio between the widths of two adjacent
cells, is 1.04. The coarse grained solid volume fraction along the horizontal center line is
presented in Fig.~\ref{fig:diff-str}(b). It can be seen that $\varepsilon_s$ profile obtained by
using the diffusion-based method on the stretched mesh are identical to that from the uniform
mesh. This demonstrates the capability of the diffusion-based method in performing averaging
on stretched meshes without causing computational artifacts.

\begin{figure}[!htpb]
  \centering
  \subfloat[][unstructured mesh]{
    \includegraphics[width=0.45\textwidth]{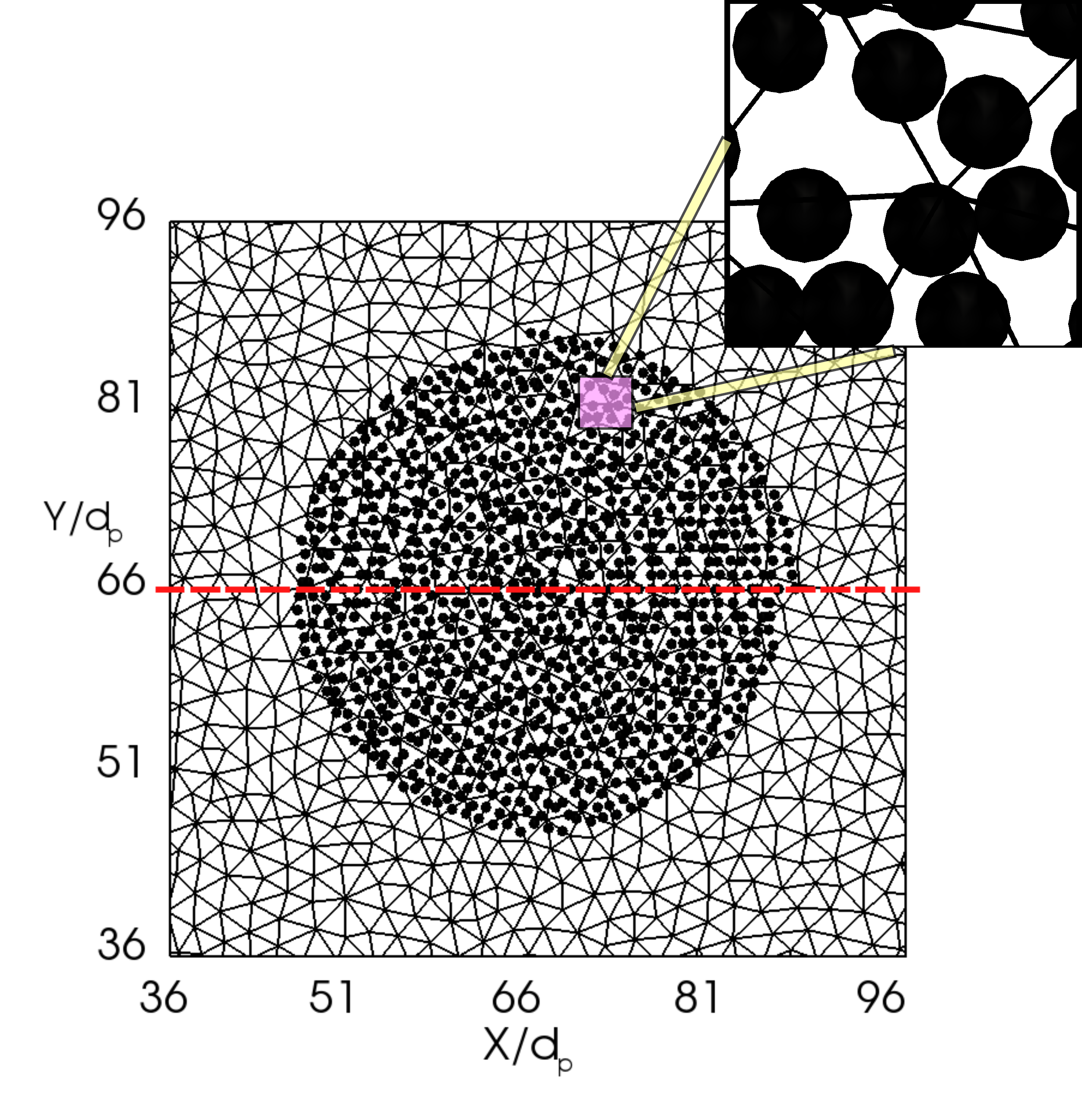}
  }
  \hspace{0.01\textwidth}
  \subfloat[][solid volume fraction]{
    \includegraphics[width=0.45\textwidth]{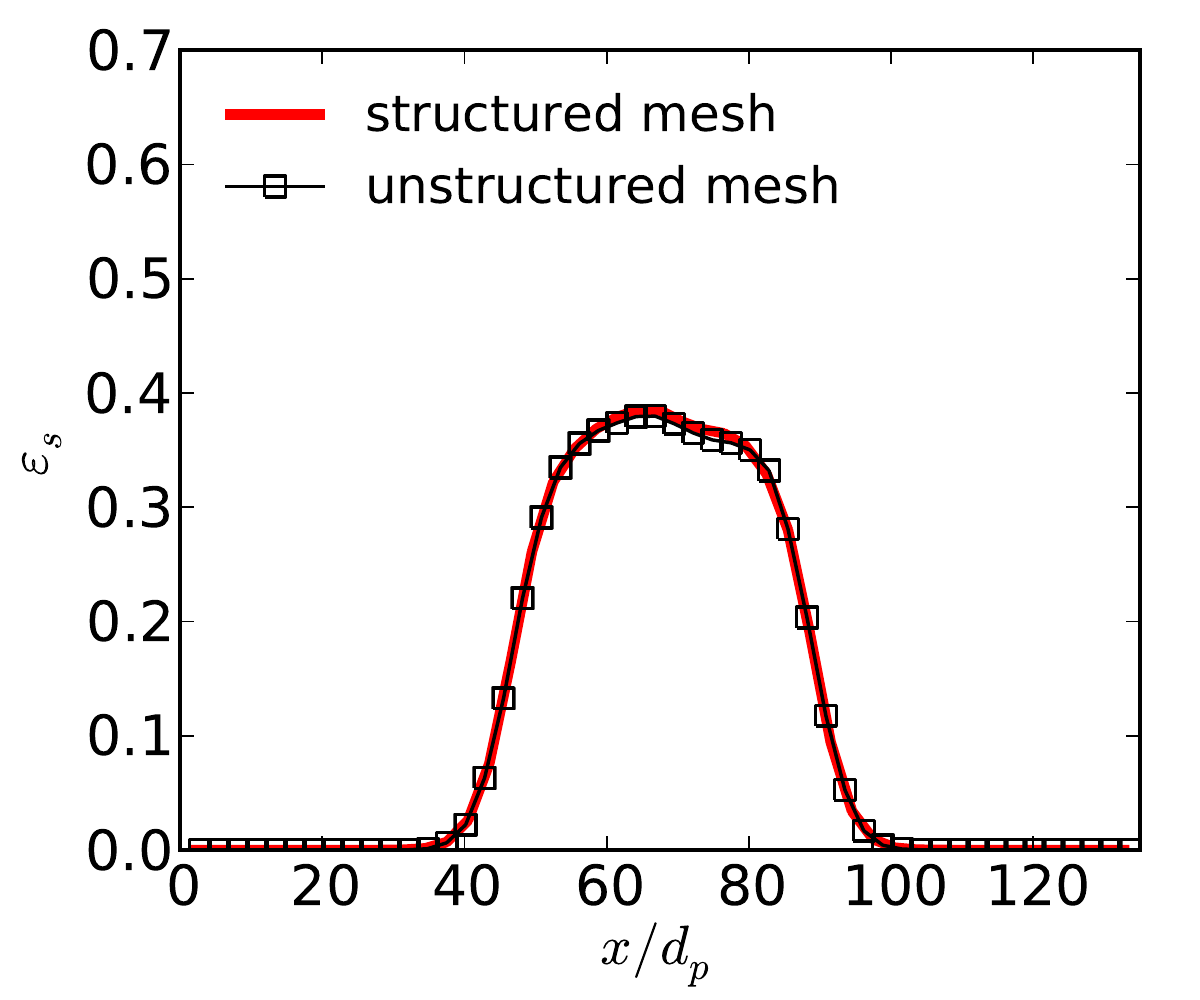}
  }
  \caption{Comparison of solid volume fractions obtained by using the diffusion-based coarse
    graining method on structured and unstructured meshes. Panel (a) displays the mesh and the
    particle distribution in the shaded region of Fig.~\ref{fig:apriori-domain}(a), with an inset
    showing a close-up view of a small region. The solid volume fraction profile along the
    horizontal centerline of the domain is shown in panel (b).  \label{fig:diff-uns} }
\end{figure}

The coarse graining test on an unstructured mesh with triangular cells are shown in
Fig.~\ref{fig:diff-uns}.  Figure~\ref{fig:diff-uns}(a) illustrates the particle distribution and the
unstructured mesh, with a close-up view of part of the computational domain shown in the inset. The
obtained solid volume fraction along the horizontal center line is shown in
Fig.~\ref{fig:diff-uns}(b). It can be seen that the coarse grained $\varepsilon_s$ profile obtained
on the unstructured mesh is consistent with that obtained on the structured mesh.

It is worth mentioning that it would be difficult for other averaging methods to obtain smooth
$\varepsilon_s$ fields for stretched or irregular meshes as the ones shown in
Figs.~\ref{fig:diff-str}(a) and~\ref{fig:diff-uns}(a). Generally speaking, {\color{black} the methods
  that directly use the CFD mesh for averaging (e.g., PCM and DPVM) are more susceptible to the
  spatial variations (e.g., stretching) in the CFD meshes. In contrast, the methods that use an
  independent mesh or parameter for averaging (e.g., two-grid formulation and statistical kernel
  method, respectively) are more robust on stretched meshes.  We note that while the proposed method
  uses the CFD mesh for averaging, the results are relatively independent of the CFD mesh. } In
summary, the two tests here suggest that the diffusion-based averaging method can produce smooth
$\varepsilon_s$ fields even on meshes with significant stretching and on unstructured meshes with
very small cells.

\subsection{Mesh Convergence Study}
\label{sec:apriori-independence}

As stated in Section~\ref{sec:wish}, a highly desirable property of a averaging algorithm is
being able to yield mesh-independence averaged fields, where the mesh here refers to CFD
mesh. Part of the mesh-independence has been studied in Section~\ref{sec:apriori-meshshape}
above. This subsection focuses on the convergence of results on progressively refined meshes. As
reviewed in Section~\ref{sec:cg-summary} (see Table~\ref{tab:pro-con} for a summary), statistical
kernel function methods and the two-grid formulation can give relatively mesh-independent results,
since the bandwidth (for the statistical kernel methods) or mesh (for the two-grid formulation) used
for averaging can be chosen independently of the CFD mesh.  In contrast, PCM and DPVM use the
CFD mesh for averaging, and thus the coarse grained fields so obtained inevitably depend on
the CFD mesh used. The purpose of this test is to demonstrate the mesh convergence characteristics
of the diffusion-based averaging method.

The particle distribution and the computational domain size are the same as above as shown in
Fig.~\ref{fig:apriori-domain}(a). The coarse grained $\varepsilon_s$ fields are obtained by using the
diffusion-based method with four successively refined meshes A, B, C, and D, the resolution and cell
sizes for which are presented in Table~\ref{tab:apriori-indep}.  Note that in the finest mesh D, the
volume of each cells is $0.25d_p^3$, which is only about half of the volume of a spherical particle
of diameter $d_p$. When the volume of a cell is smaller than the particle, and if PCM is used to
obtain the solid volume fraction, the $\varepsilon_s$ value would be larger than one for the cells
in which the particle centroids are located; for DVPM, the cells that are fully occupied by a
particle would have $\varepsilon_s$ values equal to one.  Even though such scenarios may only occur
in a few cells, it can lead to very large fluid drag forces on the particles and on the fluid,
ultimately destabilizing the simulation.

\begin{figure}[!htpb]
  \centering
\subfloat[][diffusion-based method]{
  \includegraphics[width=0.495\textwidth]{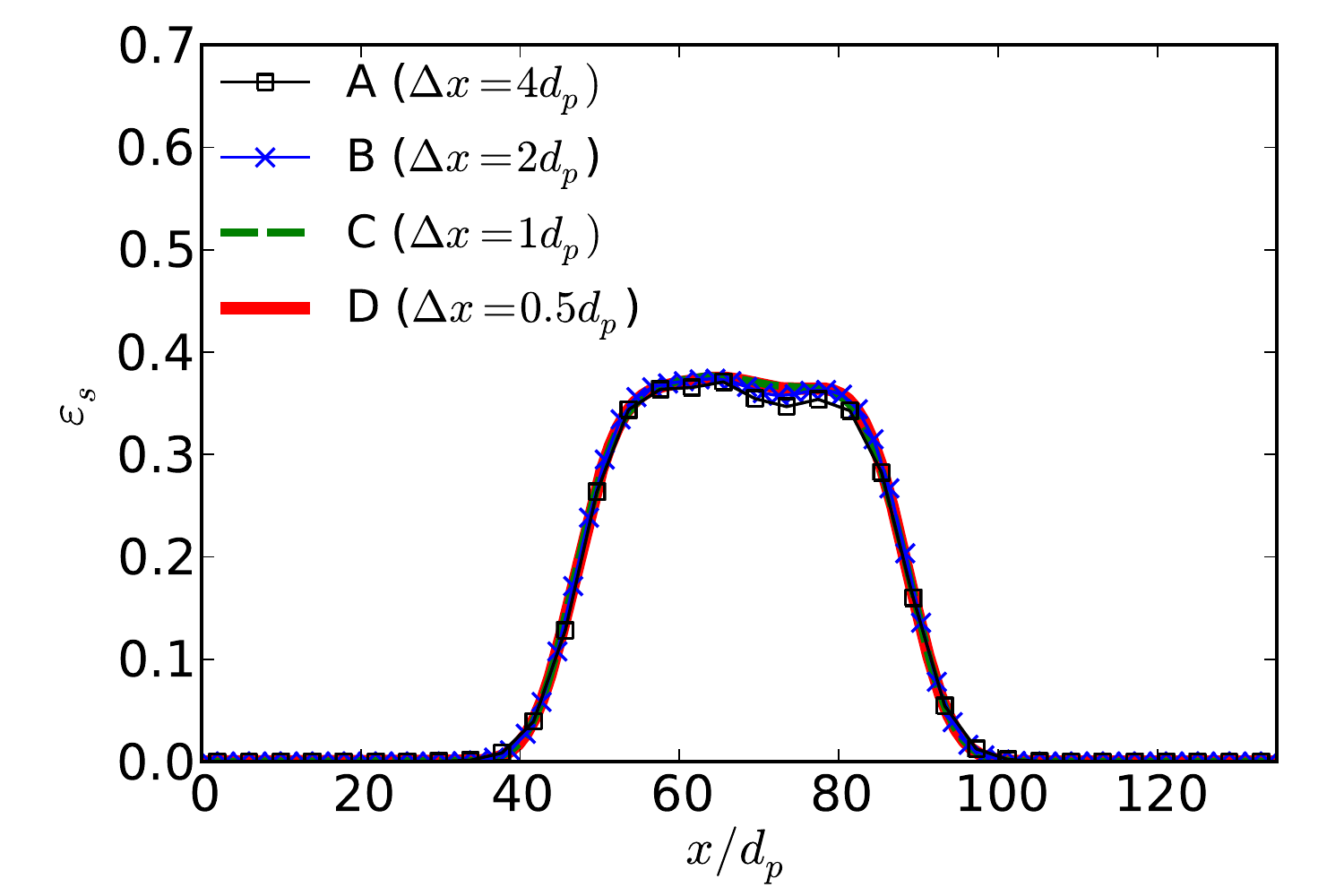}
} \\
\subfloat[][PCM]{
  \includegraphics[width=0.495\textwidth]{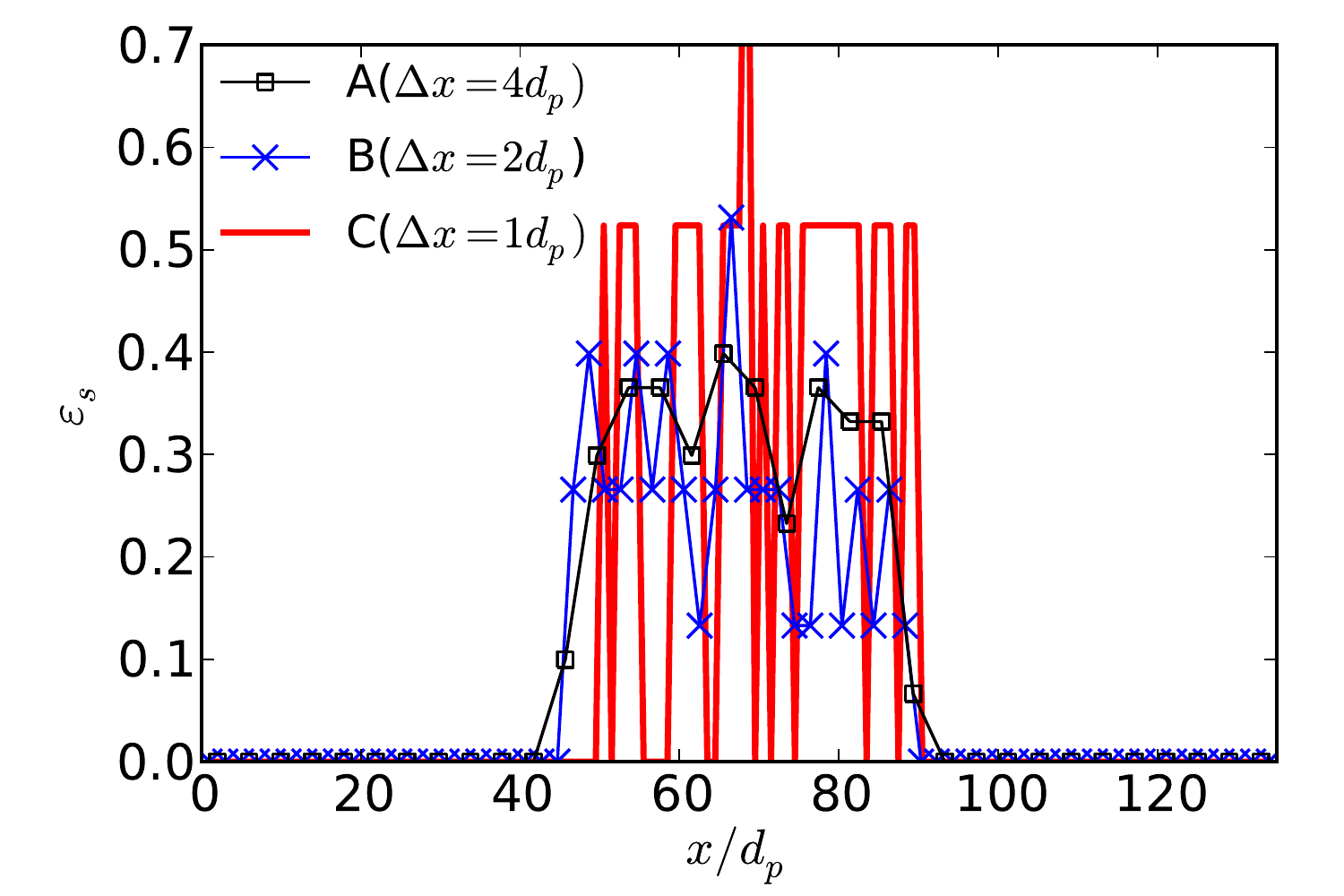}
}
\subfloat[][DPVM]{
  \includegraphics[width=0.495\textwidth]{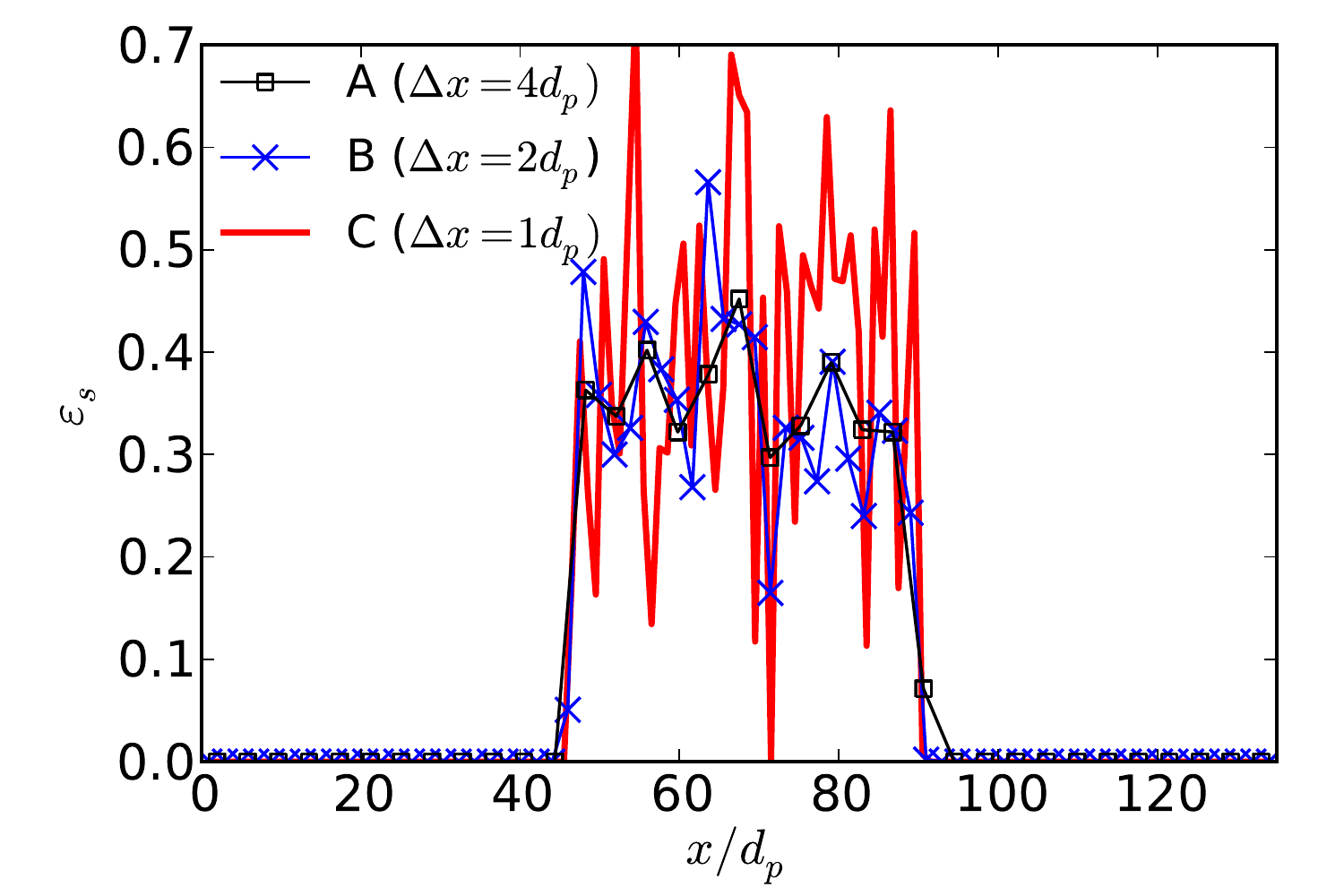}
}
\caption{
  \label{fig:diff-indep}
  Mesh independence test of the proposed method, showing the coarse grained solid volume fraction
  $\varepsilon_s$ fields along the horizontal centerline of the domain (indicated by the dashed line
  in Fig.~\ref{fig:apriori-eq1}(a)) for (a) diffusion-based method, (b) PCM, and (c) DPVM. The
  study is performed on four successively refined meshes A--D, with the cell dimensioned and the numbers
  of cells presented in Table~\ref{tab:apriori-indep}. The results for PCM and DPVM on the finest
  mesh ($\Delta x = 0.5d_p$) are omitted due to the presence of excessively large $\varepsilon_s$
  values, and also because these two methods are not expected to work on this mesh with $\Delta x$
  smaller than the particle diameter $d_p$.}
\end{figure}

The coarse grained $\varepsilon_s$ fields obtained by using the diffusion-based method on meshes
A--D are presented in Fig.~\ref{fig:diff-indep}(a).  It can be seen that the coarse grained
$\varepsilon_s$ fields along the horizontal centerline obtained with the three finer meshes B, C and
D are identical (as demonstrated by the lines falling on top of each other), suggesting that mesh
convergence is achieved. The very minor deviation of the results from mesh A ($\Delta x = 4 d_p$)
near $x/d_p = 75$ is due to the slightly inadequate mesh resolution, and this is expected. Also note
from Fig.~\ref{fig:diff-indep} that unphysically large $\varepsilon_s$ values do not appear in any
of the coarse grained $\varepsilon_s$ fields produced by the diffusion-based method, even for the
finest mesh D ($\Delta x = 0.5 d_p$), which has cell volumes smaller than the volume of a single
particle as explained above.  For comparison purposes, the coarse grained $\varepsilon_s$ fields
obtained by using PCM and DPVM are presented in Figs.~\ref{fig:diff-indep}(b) and
\ref{fig:diff-indep}(c), respectively. Only the results from meshes A--C are shown for these two
cases, and the results for mesh D are omitted. This is due to the presence of excessively large
$\varepsilon_s$ values on the mesh D results, and also because PCM and DPVM are not expected to work
on meshes with cell dimensions smaller than the particle diameter. From
Figs.~\ref{fig:diff-indep}(b) and~\ref{fig:diff-indep}(c) it can be seen that with PCM and DPVM the
coarse grained solid volume fraction values exhibit more spatial oscillations as the mesh is
refined, producing more extreme high and low values (e.g., high values over 0.7 and low values of 0
in mesh C for both cases). Mesh convergences is not achieved for either methods. Although it is not
presented here, it is expected that the two-grid formulation should be able to give mesh-independent
results, as long as the same mesh (e.g., mesh A) is used for as the coarse-graining mesh same for
the all cases. In this case, the coarse-graining mesh would contain $1 \times 1$, $2 \times 2$, $4
\times 4$, and $8 \times 8$ CFD cells, respectively, for cases A, B, C, and D.

Hence, from the results obtained in the mesh convergence test and those presented in
Section~\ref{sec:apriori-meshshape}, we can conclude that the diffusion-based method is able to
produce smooth and mesh-independent coarse grained fields, even for very fine meshes with cells that
are smaller than the particles or meshes with stretching. This is a significant advantage over PCM
and DPVM, and would have profound implications in CFD--DEM simulations.

\begin{table}[!htbp]
  \caption{Parameters for the four successively refined meshes used in the mesh-convergence study}
  \begin{center}
  \begin{tabular}{ccc}
    \hline
    mesh & $\Delta x$ and $\Delta y$ & $N_x$ and $N_y$ \\
    \hline
    A & 4$d_p$ & $35$ \\
    B & 2$d_p$ & $69$ \\
    C & 1$d_p$ & $135$ \\
    D & 0.5$d_p$ & $270$ \\
    \hline
  \end{tabular}
  \end{center}
  \label{tab:apriori-indep}
\end{table}

\section{Discussion}
\label{sec:discuss}

\subsection{Implementation and Computational Overhead}
\label{sec:overhead}
The proposed averaging method involves solving transient diffusion equations with no-flux
boundary conditions, which is straightforward and can usually take advantage of the existing
infrastructure (e.g., discretization schemes, linear system solvers, parallel computing
capabilities) available in the CFD solver. In this work, the implementation of the diffusion
equation solver in OpenFOAM and its integration into the CFD--DEM solver needed only a few dozen
lines of additional code. {\color{black} In particular, the no-flux boundary condition is easy to
  enforce without additional complexity in a finite-volume solver such as OpenFOAM. In
  finite-difference-based solvers, ghost nodes may be needed for discretization of the Laplacian
  operator. However, usually these special treatments are already in place in the CFD solver, and
  the ghost nodes are only limited to a few layers depending on the stencil width of the
  discretization scheme. This is in contrast to the ``image'' regions (see
  Fig.~\ref{fig:kernel-img}a) in statistical kernel methods, where the number of cell layers is of
  the order of $b/\Delta x$ (i.e., the ratio between bandwidth and cell size) and can be rather
  large when the cells are small.}

A second-order central scheme was used for the spatial discretization in the diffusion equation; the
Crank--Nicolson scheme was used for temporal integration. With this choice of discretization schemes
the stability is guaranteed for any time step size due to the implicit nature of the time
stepping. Therefore, the time step size is only restricted by the solution accuracy to be
achieved. It is noted that the accuracy in solving the diffusion equations is \emph{not} essential
for the coarse graining. One can consider that an inaccurate temporal or spatial discretization of
the diffusion equation leads to a solution corresponding to a modified equation, which may have a
different diffusion constant (corresponding to a different kernel bandwidth) than specified, or even
a different way of averaging.  However, slight inaccuracies are acceptable, since the coarse
graining scheme used to link microscopic and macroscopic variables is non-unique anyway. Hence, for
the sake of reducing computational costs, large time step sizes can be used to solve the diffusion
equations in the averaging.

\begin{figure}[!htpb]
  \centering
      \includegraphics[width=0.6\textwidth]{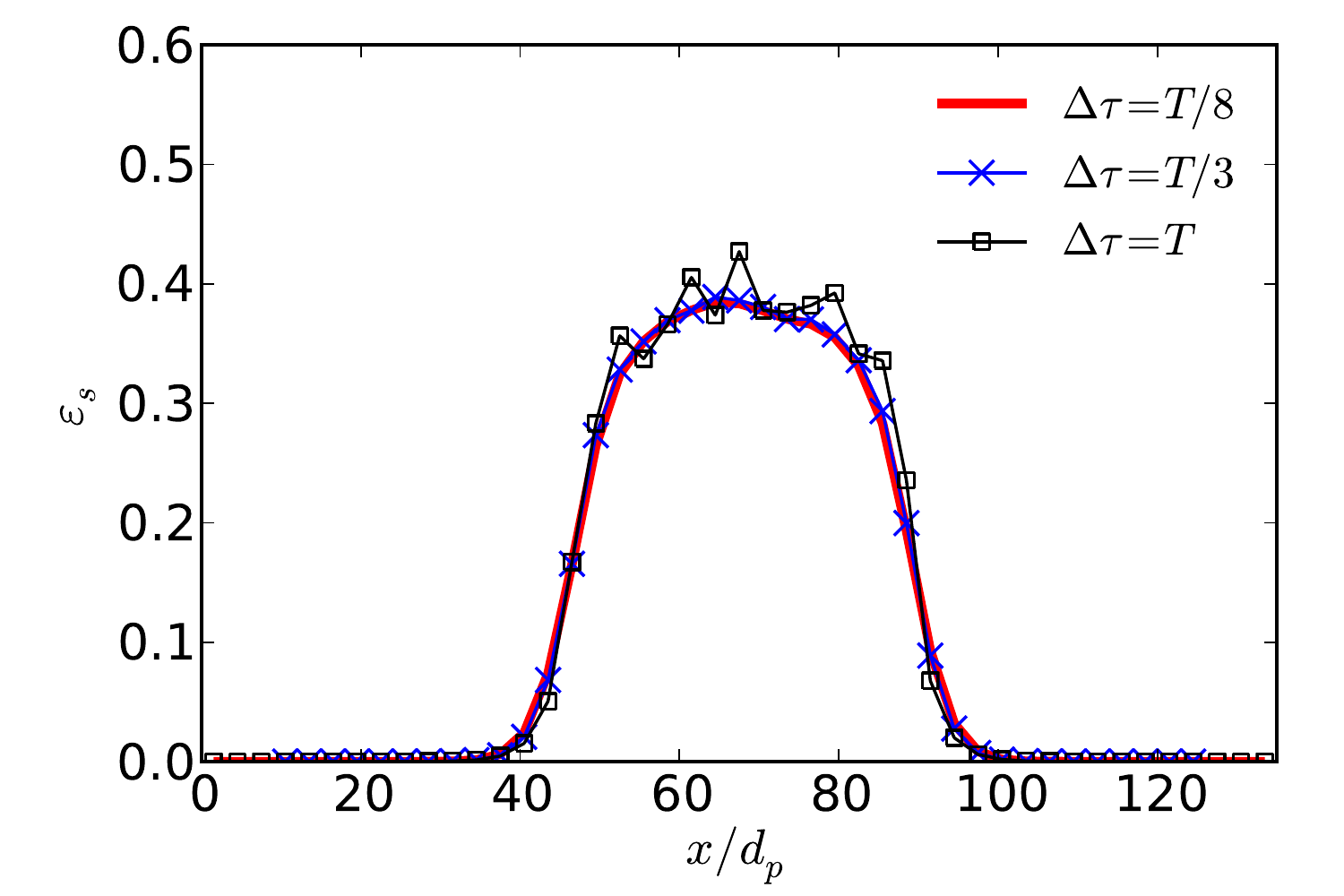}
      \caption{Effect of time step sizes used to solve the diffusion equation in the averaging procedure.
   \label{fig:diff-dt-study} }
\end{figure}

In this study it was found that using only three or even one time step, i.e., a time step size of
\(\Delta \tau = T/3\) or \(T\), was sufficient. Figure~\ref{fig:diff-dt-study} displays the coarse
grained $\varepsilon_s$ field obtained with different time step sizes $\Delta \tau$ = $T/8$, $T/3$,
and $T$. It can be seen that the \( \varepsilon_s \) fields obtained in the three case are almost
identical.  The particle configuration presented in Fig.~\ref{fig:apriori-eq1}(a) was used in this
calculation. The number of time steps needed to solve the diffusion equation with a given accuracy
also depends on the time span $T$ or the bandwidth $b$. The bandwidth used in obtaining the results
in Fig.~\ref{fig:diff-dt-study} was $b=6 d_p$. Bandwidths up to \( b=24d_p \) have been tested, and
even with such a excessively large bandwidth, a time step size of \(\Delta \tau = T/3\) or \(T\)
still yielded qualitatively similar results as presented in Fig.~\ref{fig:diff-dt-study}. The kernel
bandwidth \( b \) is a physical parameter usually specified based on the particle diameters \(d_p\).
In particular, note that the bandwidth \(b\) is chosen independent of the CFD mesh sizes.

\subsection{Numerical Diffusion and Offset on Coarse Meshes}
\label{sec:num-diffu}

 As pointed out in Section~\ref{sec:connect}, the equivalence between the
  diffusion-based coarse-gaining method and the statistical kernel-based method holds rigorously
  only theoretically, i.e., when the CFD mesh is infinitely fine. The equivalence is not exact on
  meshes consisting of cells of finite sizes. Specifically, when the cell size is large compared to
  the diffusion bandwidth, numerical diffusions occur. Taking the calculation of solid volume
  fraction for example, if a particle resides in a relatively large cell ($b/\Delta x < 1$), the
  statistical kernel method assigns most of the particle volume to the host cell, but the diffusion-based
  coarse-graining method still assigns some volume to the neighboring cells via diffusion.  This
  error is closely related to the ratio $b/\Delta x$ between the diffusion bandwidth and the cell
  size.

\begin{figure}[!htpb]
    \centering \includegraphics[width=0.9\textwidth]{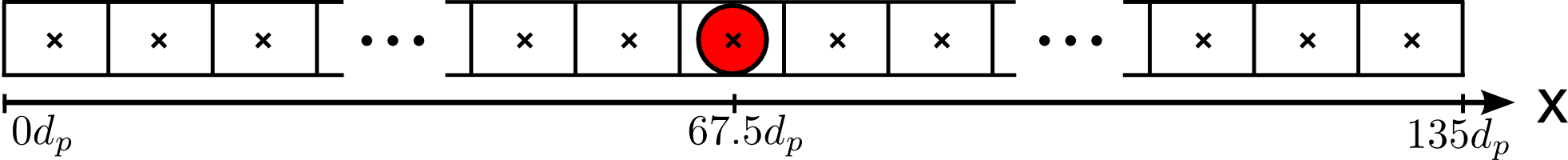}
  \caption{ Computational domain and setup in the investigation of numerical diffusion of the
  proposed method.  The one-dimensional domain has a length of $135d_p$, and a particle is located
  at $x=67.5 d_p$. Cell sizes investigated include $\Delta x = 1.5 d_p$, $3d_p$, $6d_p$, and $12
  d_p$.}
  \label{fig:diffu-schematic}
\end{figure}

\begin{figure}[!htpb]
  \centering
  \subfloat[][$\Delta x = 1.5d_p,~b/\Delta x=4$]{
    \includegraphics[width=0.45\textwidth]{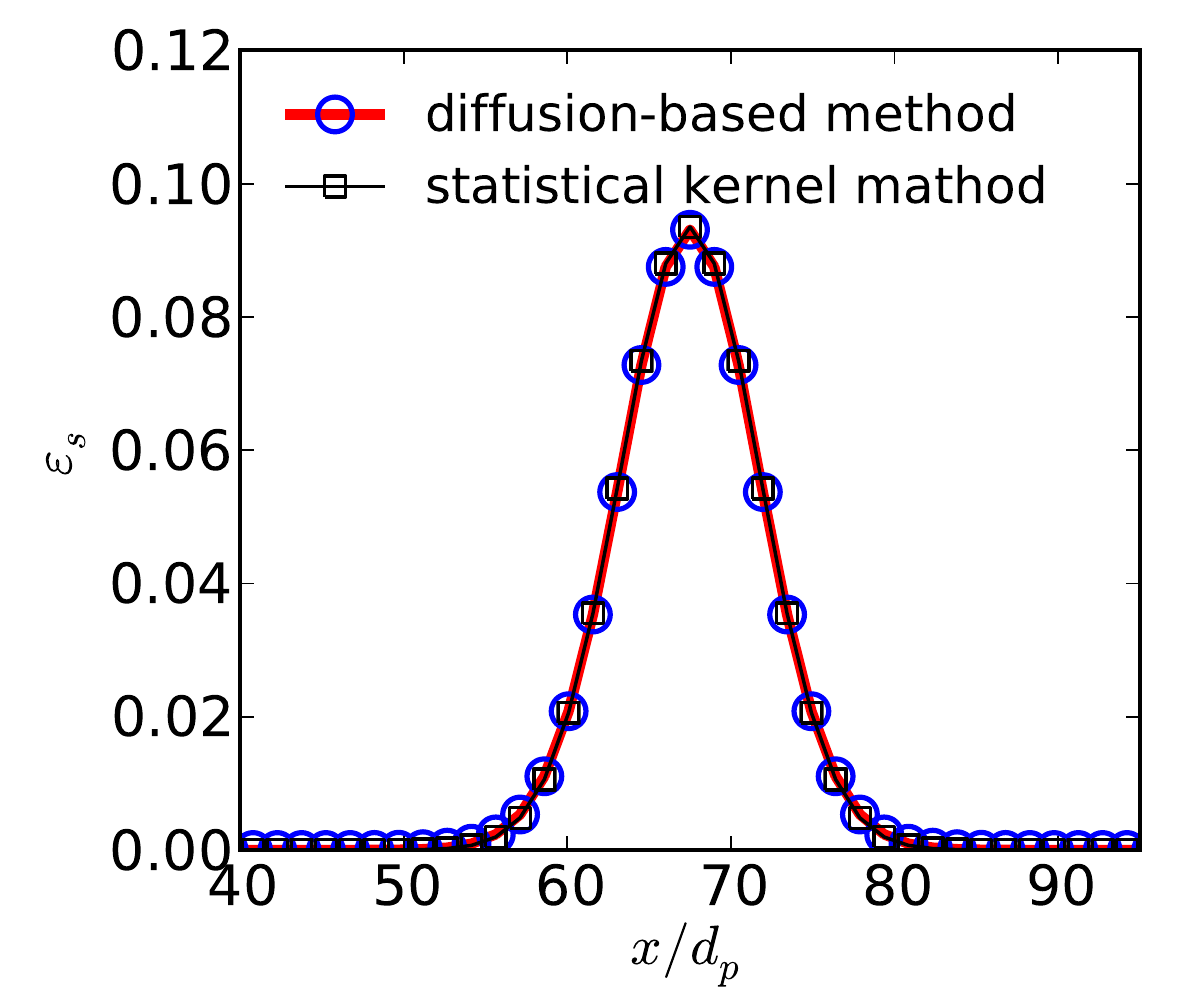}
  }
  \subfloat[][$\Delta x = 3d_p,~b/\Delta x = 2$]{
    \includegraphics[width=0.45\textwidth]{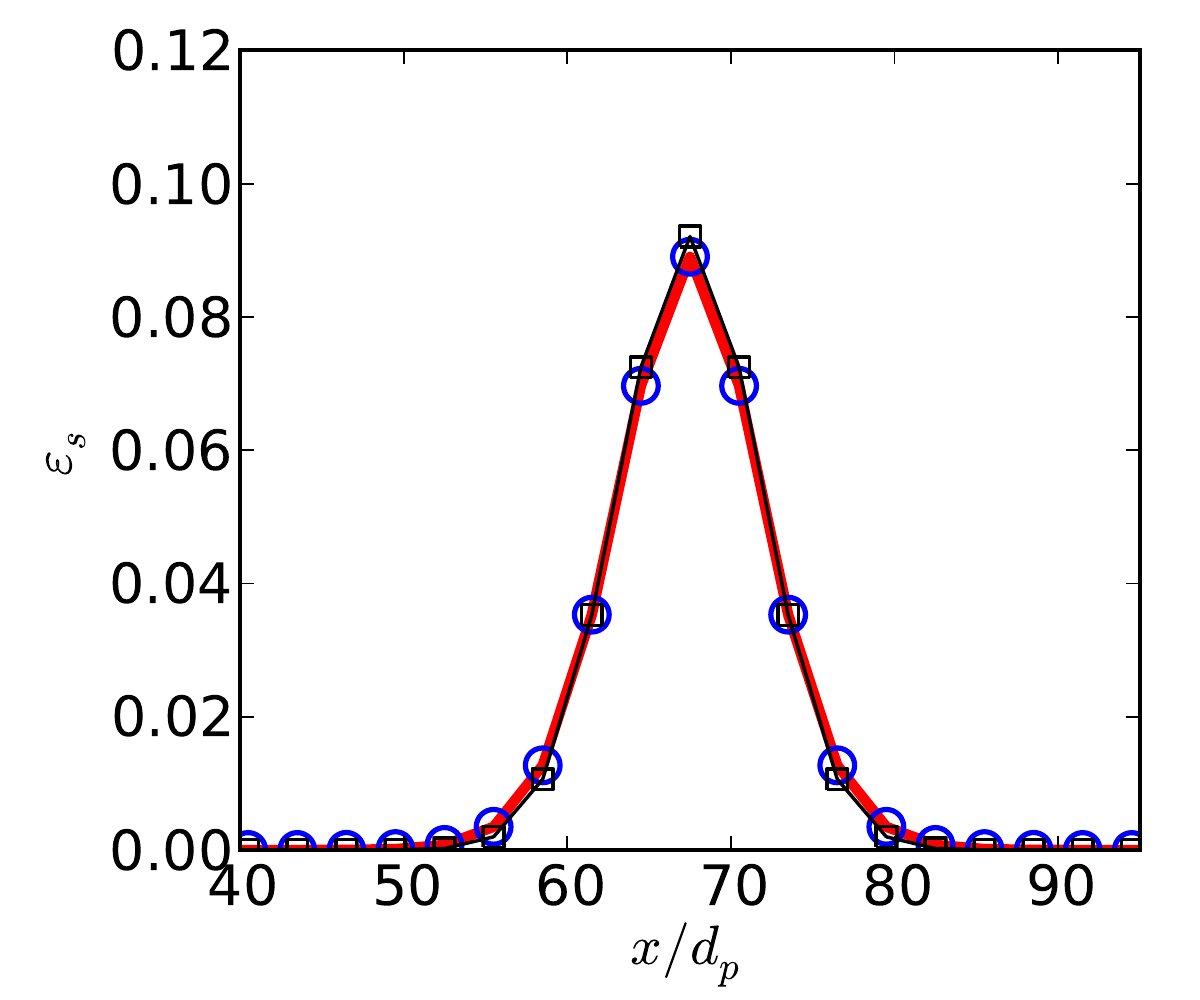}
  }
  \vspace{0.01\textwidth}
  \subfloat[][$\Delta x = 6d_p,~b/\Delta x = 1$]{
    \includegraphics[width=0.45\textwidth]{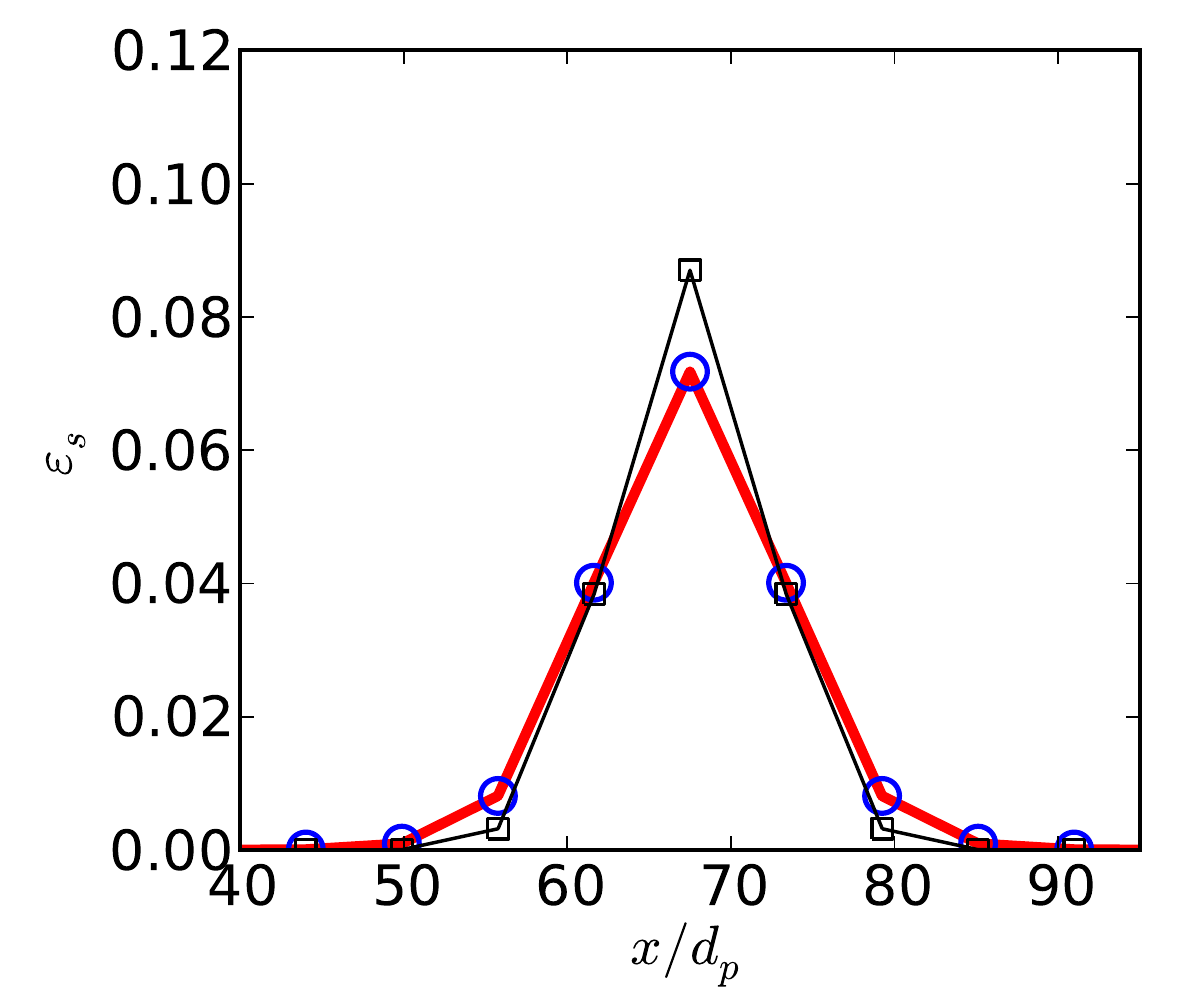}
  }
  \subfloat[][$\Delta x = 12d_p,~b/\Delta x = 0.5$]{
    \includegraphics[width=0.45\textwidth]{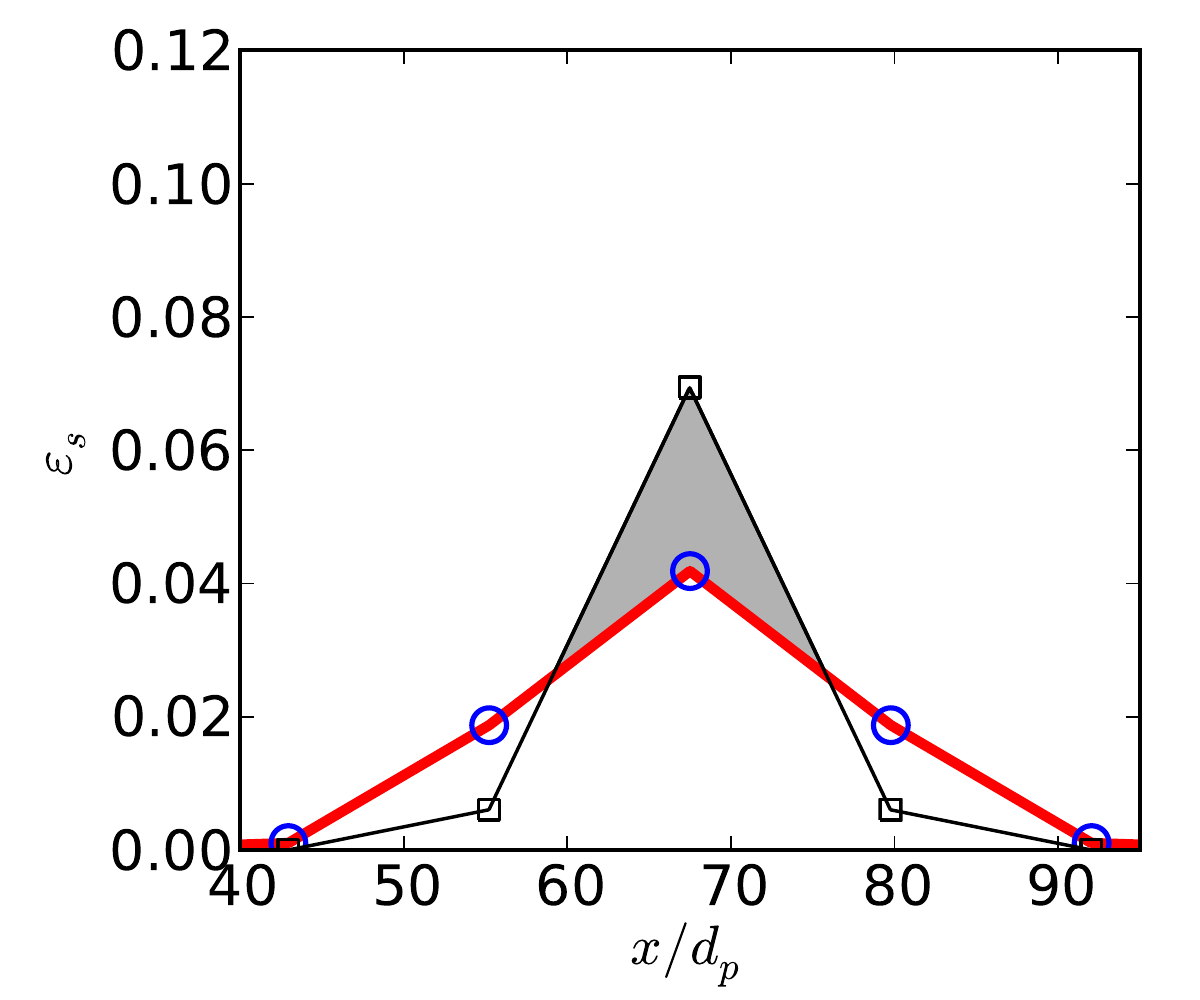}
  }
  \caption{Study of numerical diffusion due to meshes with finite-size cells. The solid volume
    fractions computed by using the diffusion-based method are compared with those obtained with the
    statistical kernel method, both with bandwidth $b=6d_p$, for four different cell sizes: (a)
    $\Delta x = 1.5 d_p$, (b) $\Delta x = 3 d_p$, (c) $\Delta x = 6d_p$, and (d) $\Delta x = 12
    d_p$, corresponding to bandwidth/cell size ratios of $b/\Delta x = 4, 2, 1$ and $0.5$,
    respectively. The shaded area in panel (d), normalized by the total area under the curve
    corresponding to the statistical kernel method, is used to indicate the extent of numerical
    diffusion.}
  \label{fig:bandwidth-dx}
\end{figure}

  To study the numerical diffusion, we use a one-dimensional domain of length $135 d_p$. The
  dimensions of the domain in height and transverse directions are both $d_p$. A particle is
  co-located with the center of a cell at $x=67.5 d_p$, as depicted in
  Fig.~\ref{fig:diffu-schematic}.  Solid volume fractions are computed with the
  diffusion-based method and the statistical kernel method, the latter of which is obtained via
  analytical evaluations of the definite integral of the kernel in each cell and is thus considered
  as benchmark.  Consistent with the setup in the test cases examined in Section~\ref{sec:apriori},
  the diffusion bandwidth $b$ is chosen to be $6 d_p$ for both the diffusion-based method and the
  statistical-kernel method. Four cell sizes $\Delta x = 1.5 d_p, 3d_p, 6d_p$ and $12 d_p$ are
  investigated, corresponding to bandwidth/cell size ratios of $b/\Delta x = 4$, $2$, $1$, and
  $0.5$, respectively.  The obtained solid volume fractions are presented in
  Fig.~\ref{fig:bandwidth-dx}.

It can be seen that the discrepancy between the diffusion-based method and the statistical kernel
method, which can be considered as numerical diffusion, are almost negligible for $b/\Delta x = 4$
and $2$.  The numerical diffusion is appreciable for $b/\Delta x = 1$ ($\Delta x = 6 d_p$), but this
level of diffusion is likely to be acceptable for most applications. For $b/\Delta x = 0.5$, the
numerical diffusion is significant and is probably unacceptable.  In practice, however, such a small
bandwidth (or large cell size) is uncommon.

To further quantify the amount of numerical diffusion, we define a metric $\gamma =
A_{\textrm{diffu}}/A_{\textrm{total}}$, where $A_{\textrm{diffu}}$ is the area between the analytical
solution (from the statistical kernel method) and the numerical solution (from the diffusion-based
method) around the peak, as indicated by the shaded area in Fig.~\ref{fig:bandwidth-dx}(d), and
$A_{\textrm{total}}$ is the total area under the analytical solution curve. If the numerical
solution is identical to the analytical solution, then $\gamma = 0$; if the numerical solution is completely
diffused to the entire domain, then $\gamma$ is approximately $100\%$ for a domain that is very large (compared
with the diffusion bandwidth). The results for the four cases presented in
Fig.~\ref{fig:bandwidth-dx} are evaluated with the metric defined above, and we obtained $\gamma =
0.6\%, 2.6\%, 8.1\%$, and $23.1\%$, respectively, for $b/\Delta x = 4, 2, 1$, and $0.5$. A correlation between
$\gamma$ and $b/\Delta x$ can be established from this investigation. In practical simulations, the
ratio $b/\Delta x$ is known in the entire field, which can then be used to infer $\gamma$ and to
assess the possible extent of numerical diffusion before actual simulations are performed.  If
numerical diffusion is a concern, in regions with large cells one can simply degenerate to PCM ,
which is particularly suitable for large cells. Within the proposed framework, the degeneration can
be straightforwardly achieved by locally setting the diffusion constant (discussed above in
Section~\ref{sec:overhead}) to very small values in the regions of concern. Conservation would still
be guaranteed in the diffusion-based averaging.

Another limitation of the proposed coarse-graining method is that it does not distinguish the specific
locations of the particles within its host cell. Instead, all particle volumes are first lumped to
the host cell center as in the PCM, and then the obtained field is used as initial condition to solve the
diffusion equation. This limitation is directly inherited from the PCM. Potentially, the obtained
volume fraction distribution may have an offset error, and the amount of offset for a particular
particle can be as large as the distance from the furtherest point in the host cell to the cell
center. However, when there are a large number of particles randomly distributed in the
domain, the offset errors associated with individual particles tend to cancel. This is rarely a
concern in most simulations, and it decreases with mesh refinement.

\section{Conclusion}
\label{sec:conclude}
In this work, we proposed a coarse-graining algorithm based on solving diffusion equations with
no-flux boundary conditions. The coarse graining method is valuable for coupled continuum--discrete
solvers such as CFD--DEM solvers for dense particle-laden flows.  The proposed algorithm can be
straightforwardly implemented in any parallel, three-dimensional mesh-based CFD solvers developed
for industrial flow simulations on complex geometries.  Via theoretical analysis we demonstrated
that the proposed method was equivalent to the statistical Gaussian kernel-based coarse graining
method. The equivalence is established for both interior particles and particles near boundaries,
and is valid for domains of arbitrary domain shapes up to the mesh discretization accuracy. The
theoretical equivalence was verified by using several \textit{a priori} numerical tests. It was further
demonstrated that the proposed algorithm was able to yield physically reasonable, spatially smooth
coarse grained fields on stretched meshes, unstructured meshes, and meshes having cells with volumes
smaller than the volume of a particle. Moreover, numerical tests on successively refined meshes
showed that the diffusion-based coarse graining method was able to produced mesh-independent
results, which is an important advantage over many existing coarse graining methods such as the
particle centroid method and the divided particle volume method.

In summary, the merits of the proposed coarse graining method include (1) sound theoretical
foundation, (2) unified treatment of interior and near-boundary particles within the same framework,
(3) guaranteed conservation of relevant physical quantities (e.g., particle mass in the cases demonstrated) in
the coarse graining procedure, (4) easy implementation in CFD solvers with almost arbitrary meshes
and ability to produce smooth and mesh-independent coarse-grained fields on unfavorable meshes, and
(5) easy parallelization by utilizing the existing infrastructure in the CFD solver. 

 The diffusion-based coarse graining method has been implemented into a CFD--DEM
  solver. We emphasize that caution should be exercised when using this method in CFD--DEM
  solvers. In particular, the diffusion should be applied on the momentum and not on the velocities.
  Detailed derivations and CFD--DEM simulation results obtained with the coarse graining method
  proposed here are presented in \citep{Xiao-IJMF}.

\section{Acknowledgment}

The computational resources used for this project were provided by the Advanced Research Computing
(ARC) of Virginia Tech, which is gratefully acknowledged.  Authors gratefully acknowledge partial
funding of graduate research assistantship from the Institute for Critical Technology and Applied
Science (ICTAS, grant number 175258) in this effort.  We thank the Dr. Chunliang Wu of ANSYS for his
comments, which helped improving the quality of the manuscript.

\bibliographystyle{elsarticle-harv}

\end{document}